\documentclass[twocolumn]{aastex63}
\usepackage{url}

\usepackage{natbib}
\usepackage{appendix}
\received{}
\revised{}
\accepted{May 26, 2021}
\submitjournal{ApJ in Press}

\shorttitle{Morphology, Stellar Population, and Dynamics of EMPGs}
\shortauthors{Isobe et al.}
\graphicspath{{./}{figures/}}
\def\tcr{\textcolor{black}}
\def\tcra{\textcolor{black}}
\def\tcg{\textcolor{black}}
\begin{document}

\title{EMPRESS. III.\\
Morphology, Stellar Population, and Dynamics of Extremely Metal Poor Galaxies (EMPGs):\\
Are EMPGs Local Analogs of High-$z$ Young Galaxies?\footnote{Released on}}

\author[0000-0001-7730-8634]{Yuki Isobe}
\affiliation{Institute for Cosmic Ray Research, The University of Tokyo, 5-1-5 Kashiwanoha, Kashiwa, Chiba 277-8582, Japan}
\affiliation{Department of Physics, Graduate School of Science, The University of Tokyo, 7-3-1 Hongo, Bunkyo, Tokyo 113-0033, Japan}

\author[0000-0002-1049-6658]{Masami Ouchi}
\affiliation{National Astronomical Observatory of Japan, 2-21-1 Osawa, Mitaka, Tokyo 181-8588, Japan}
\affiliation{Institute for Cosmic Ray Research, The University of Tokyo, 5-1-5 Kashiwanoha, Kashiwa, Chiba 277-8582, Japan}
\affiliation{Kavli Institute for the Physics and Mathematics of the Universe (WPI), University of Tokyo, Kashiwa, Chiba 277-8583, Japan}

\author[0000-0001-5780-1886]{Takashi Kojima}
\affiliation{Institute for Cosmic Ray Research, The University of Tokyo, 5-1-5 Kashiwanoha, Kashiwa, Chiba 277-8582, Japan}
\affiliation{Department of Physics, Graduate School of Science, The University of Tokyo, 7-3-1 Hongo, Bunkyo, Tokyo 113-0033, Japan}

\author{Takatoshi Shibuya}
\affiliation{Kitami Institute of Technology, 165 Koen-cho, Kitami, Hokkaido 090-8507, Japan}

\author{Kohei Hayashi}
\affiliation{Institute for Cosmic Ray Research, The University of Tokyo, 5-1-5 Kashiwanoha, Kashiwa, Chiba 277-8582, Japan}

\author{Michael Rauch}
\affiliation{Carnegie Observatories, 813 Santa Barbara Street, Pasadena, CA 91101, USA}

\author[0000-0003-2449-6314]{Shotaro Kikuchihara}
\affiliation{Institute for Cosmic Ray Research, The University of Tokyo, 5-1-5 Kashiwanoha, Kashiwa, Chiba 277-8582, Japan}
\affiliation{Department of Astronomy, Graduate School of Science, The University of Tokyo, 7-3-1 Hongo, Bunkyo, Tokyo 113-0033, Japan}

\author{Haibin Zhang}
\affiliation{Institute for Cosmic Ray Research, The University of Tokyo, 5-1-5 Kashiwanoha, Kashiwa, Chiba 277-8582, Japan}

\author[0000-0001-9011-7605]{Yoshiaki Ono}
\affiliation{Institute for Cosmic Ray Research, The University of Tokyo, 5-1-5 Kashiwanoha, Kashiwa, Chiba 277-8582, Japan}

\author[0000-0001-7201-5066]{Seiji Fujimoto} 
\affiliation{Cosmic DAWN Center}
\affiliation{Niels Bohr Institute, University of Copenhagen, Lyngbyvej2, DK-2100, Copenhagen, Denmark}
\affiliation{Research Institute for Science and Engineering, Waseda University, 3-4-1 Okubo, Shinjuku, Tokyo 169-8555, Japan}
\affiliation{National Astronomical Observatory of Japan, 2-21-1 Osawa, Mitaka, Tokyo 181-8588, Japan}
\affiliation{Institute for Cosmic Ray Research, The University of Tokyo, 5-1-5 Kashiwanoha, Kashiwa, Chiba 277-8582, Japan}

\author[0000-0002-6047-430X]{Yuichi Harikane} 
\affiliation{National Astronomical Observatory of Japan, 2-21-1 Osawa, Mitaka, Tokyo 181-8588, Japan}
\affiliation{Department of Physics and Astronomy, University College London, Gower Street, London WC1E 6BT, UK}

\author[0000-0002-1418-3309]{Ji Hoon Kim}
\affiliation{Subaru Telescope, National Astronomical Observatory of Japan, National Institutes of Natural Sciences (NINS), 650 North Aohoku Place, Hilo, HI 96720, USA}
\affiliation{Metaspace, 36 Nonhyeon-ro, Gangnam-gu, Seoul 06312, Republic of Korea}

\author{Yutaka Komiyama} 
\affiliation{National Astronomical Observatory of Japan, 2-21-1 Osawa, Mitaka, Tokyo 181-8588, Japan}

\author[0000-0002-3801-434X]{Haruka Kusakabe} 
\affiliation{Observatoire de Gen{\'e}ve, Universit{\'e} de Gen{\'e}ve, 51 Ch. des Maillettes, 1290 Versoix, Switzerland}

\author[0000-0003-1700-5740]{Chien-Hsiu Lee} 
\affiliation{Subaru Telescope, National Astronomical Observatory of Japan, National Institutes of Natural Sciences (NINS), 650 North Aohoku Place, Hilo, HI 96720, USA}

\author[0000-0003-4985-0201]{Ken Mawatari}
\affiliation{Institute for Cosmic Ray Research, The University of Tokyo, 5-1-5 Kashiwanoha, Kashiwa, Chiba 277-8582, Japan}

\author[0000-0003-3228-7264]{Masato Onodera} 
\affiliation{Subaru Telescope, National Astronomical Observatory of Japan, National Institutes of Natural Sciences (NINS), 650 North Aohoku Place, Hilo, HI 96720, USA}
\affiliation{Department of Astronomical Science, SOKENDAI (The Graduate University for Advanced Studies), Osawa 2-21-1, Mitaka, Tokyo, 181-8588, Japan}

\author[0000-0001-6958-7856]{Yuma Sugahara} 
\affiliation{National Astronomical Observatory of Japan, 2-21-1 Osawa, Mitaka, Tokyo 181-8588, Japan}
\affiliation{Waseda Research Institute for Science and Engineering, Faculty of Science and Engineering, Waseda University, 3-4-1, Okubo, Shinjuku, Tokyo 169-8555, Japan}
\affiliation{Institute for Cosmic Ray Research, The University of Tokyo, 5-1-5 Kashiwanoha, Kashiwa, Chiba 277-8582, Japan}
\affiliation{Department of Physics, Graduate School of Science, The University of Tokyo, 7-3-1 Hongo, Bunkyo, Tokyo 113-0033, Japan}

\author[0000-0001-6229-4858]{Kiyoto Yabe}
\affiliation{Kavli Institute for the Physics and Mathematics of the Universe (WPI), University of Tokyo, Kashiwa, Chiba 277-8583, Japan}





\begin{abstract}
We present the morphology and stellar population of 27 
\tcg{extremely metal poor galaxies (EMPGs) \tcg{at $z\sim 0$ with metallicities of 0.01--0.1 Z$_{\odot}$}.
}
We conduct multi-component surface brightness (SB) profile fitting for the deep Subaru/HSC $i$-band images of the EMPGs with the {\sc Galfit} software, carefully removing the SB contributions of 
\tcr{tails}. We find that the EMPGs
with a median stellar mass of \tcra{$\log(M_{*}/{\rm M}_{\odot})=6.0$} have a median S{\'e}rsic index of $n=1.1$ and a median effective radius of \tcr{$r_{\rm e}=200$} pc, suggesting that typical EMPGs \tcr{have} very compact disk. 
We compare \tcg{the EMPGs} with $z\sim 6$ galaxies and local galaxies on the size-mass ($r_{\rm e}$--$M_*$) diagram, and identify that 
the majority of the EMPGs have a $r_{\rm e}$--$M_*$ relation similar to $z\sim0$ star-forming galaxies rather than $z\sim6$ galaxies. Not every EMPG is a local analog of high-$z$ young galaxies in the $r_{\rm e}$--$M_*$ relation. 
A spectrum of one \tcr{pair of EMPG and tail}, so far available, indicates that the \tcr{tail} is dynamically related to the EMPG with a median velocity difference of \tcr{$\Delta V=101\pm32$ km s$^{-1}$}. This moderately-large $\Delta V$ cannot be explained by the dynamics of the \tcr{tail}, but 
likely by the infall on the \tcr{tail}. 
For the first time, we may identify the metal-poor star-forming system just now infalling into the tail.
\end{abstract}

\keywords{galaxies: formation --- 
galaxies: structure --- galaxies: star formation --- galaxies: dwarf --- galaxies: kinematics and dynamics}


\section{Introduction} \label{sec:intro}

\tcr{First galaxies form from primordial gas, producing first stars. Subsequently, supernova explosions take place, and the first galaxies quickly evolve into low-mass metal-poor galaxies \citep{Wise2012}.} 
\tcra{
Although galaxy formation studies have observed galaxies up to $z\sim 6-10$, high-$z$ galaxies, identified so far, are mostly limited to high stellar mass ($M_{*}\sim 10^{8}-10^{9}\ {\rm M}_\odot$) bright galaxies that are not primordial but matured systems \citep{Hashimoto2018}. 
Even with the forthcoming James Webb Space Telescope (JWST), it is difficult to detect high-$z$ galaxies with $M_{*}\lesssim10^6\ {\rm M}_{\odot}$ without gravitational lensing (Isobe et al. in prep.). 
Complimenting these high-$z$ galaxy observations, various studies actively investigate local young dwarf galaxies \citep[e.g.,][]{Berg2019,Izotov2021b}.
}

Among local young galaxies, EMPGs are defined as galaxies with metallicities less than $12+\log(\rm O/H)=7.69$ (e.g., \citealt{Kunth2000}; \citealt{Izotov2012}; \citealt{Guseva2017}), which corresponds to 10\% of the solar metallicity of $12+\log(\rm O/H)=8.69$ \citep{Asplund2021}. By observing local EMPGs, we can \tcg{probe star-formation activities} in the metal-deficient environment. Although EMPGs become rarer \tcg{toward lower redshifts} ($<0.2$\% of all galaxies at $z\sim0$; \citealt{Morales-Luis2011}), recent studies show the presence of EMPGs in the local universe such as J0811+4730 \citep{Izotov2018}, SBS0335-052 \citep{Izotov2009}, AGC198691 \citep{Hirschauer2016}, J1234+3901 \citep{Izotov2019}, Little Cub \citep{Hsyu2017}, DDO68 \citep{Pustilnik2005}, IZw18 \citep{Izotov1998}, and Leo P \citep{Skillman2013}. \tcg{These local EMPGs show low stellar masses \tcg{($M_{*}\sim10^{6}$--$10^{9}$ M$_{\odot}$)} and high specific star-formation rates \tcr{(sSFRs) that are defined by the ratios of star-formation rate (SFR) to $M_{*}$} (${\rm sSFR}\sim10$--100 Gyr$^{-1}$), which are similar to those of high-$z$ young galaxies (\citealt{Christensen2012a}; \citealt{Christensen2012}; \citealt{Stark2014}; \citealt{Stark2015}; \citealt{Vanzella2017}; \citealt{Mainali2017}).} Thus, EMPGs are sometimes expected to be local analogs of high-$z$ young galaxies.

\tcr{Recent studies investigate morphologies and chemical distributions of EMPGs.}
\citeauthor{Almeida2016} (\citeyear{Almeida2016}; hereafter S16) have reported that 
\tcr{more than half (57\%) of EMPGs are tadpole galaxies, each of which has a large star-forming low-metallicity clump at one end and a long diffuse structure (a.k.a. tail). \citet{SanchezAlmeida2015} show that the clumps have metallicities $\sim1$ dex lower than those of the tails, which imply that star formation in the clumps is triggered by metal-poor gas inflow.
If the scenario is true, there possibly exists a low-metallicity clump just now infalling on a diffuse galaxy that appears as a tail. 
While all of the previously known tails have smooth dynamical transitions to the clumps (i.e., all of the metal-poor clumps previously known are H {\sc ii} regions of the tails; \citealt{SanchezAlmeida2013}; \citealt{Olmo-Garcia2017}), now we need to understand whether some metal-poor clumps are individual systems separated from the tails.
}

\tcr{When we compare the EMPGs with high-$z$ galaxies, we need to consider which structure of high-$z$ galaxies we see. \citet{Bouwens2017} and \citet{kikuchihara2020} report that $z=6-8$ galaxies with stellar masses of $M_{*}=10^{6}-10^{9}\:{\rm M}_{\odot}$ have effective radii of $r_{\rm e}\sim100\:{\rm pc}$. 
However, \citet{Ma2018} predict that observed sizes of some high-$z$ galaxies are dominated by high-SB regions such as young stellar clumps because diffuse regions are easily missed. 
Thus, $r_{\rm e}$ and $M_{*}$ of high-$z$ galaxies reported by \citet{Bouwens2017} and \citet{kikuchihara2020} do not possibly represent those of the whole galaxies but those of the young stellar clumps. 
As well as high-$z$ galaxies, when we conduct shallow observations for EMPGs, it is possible that we miss the tails. 
In this sense, it may be a proper way to compare physical properties (e.g., size, mass) of high-$z$ galaxies with those of high-SB regions in the EMPGs.
}

In this study, we use the EMPG sample made by \citeauthor{Kojima2020} (\citeyear{Kojima2020}; hereafter Paper I) with \tcr{Subaru/Hyper Supreme-Cam (HSC)}. Paper I selects EMPGs from \tcr{HSC} Subaru Strategic Program (HSC-SSP) data (\citealt{Aihara2019}; \tcg{HSC instruments: \citealt{Miyazaki2018}; CCD camera: \citealt{Komiyama2018}; filter: \citealt{Kawanomoto2018}; QA system: \citealt{Furusawa2018})}. \tcg{We also utilize the EMPGs reported by S16 crossmatched with the HSC-SSP catalog.} Because HSC-SSP data are advantageous in terms of deep photometry (\tcg{5$\sigma$} $i_{\rm limit}\sim26$ mag) and good seeing size (${\rm FWHM}\sim0.6$ arcsec in the $i$-band; \citealt{Aihara2019}), we can precisely measure the size of the EMPGs and the tails. 

This paper is organized as follows. We present the EMPG sample (Paper I) in Section \ref{sec:data}. We describe \tcr{the data analysis} in Section \ref{sec:analysis}. The results are shown in Section \ref{sec:results}. We discuss the nature of EMPGs in Section \ref{sec:discuss}. Section \ref{sec:sum} summarizes our findings. Throughout this paper, magnitudes are in the AB system \tcg{\citep{Oke1983}}, and we assume a standard $\Lambda$CDM cosmology with parameters of ($\Omega_{\rm m}$, $\Omega_{\rm \Lambda}$, $H_{0}$) = (0.3, 0.7, 70 km ${\rm s}^{-1}$ ${\rm Mpc}^{-1}$). In this cosmology, an angular dimension of 1.0 arcsec corresponds to a physical length of \tcr{601} pc at \tcr{$z=0.03$}. The definition of solar metallicity ${\rm Z}_{\odot}$ is given by 12+log(O/H)=8.69 \citep{Asplund2021}.

\section{Data and Samples} \label{sec:data}
\begin{figure*}[h]
    \centering
    \includegraphics[width=14.0cm]{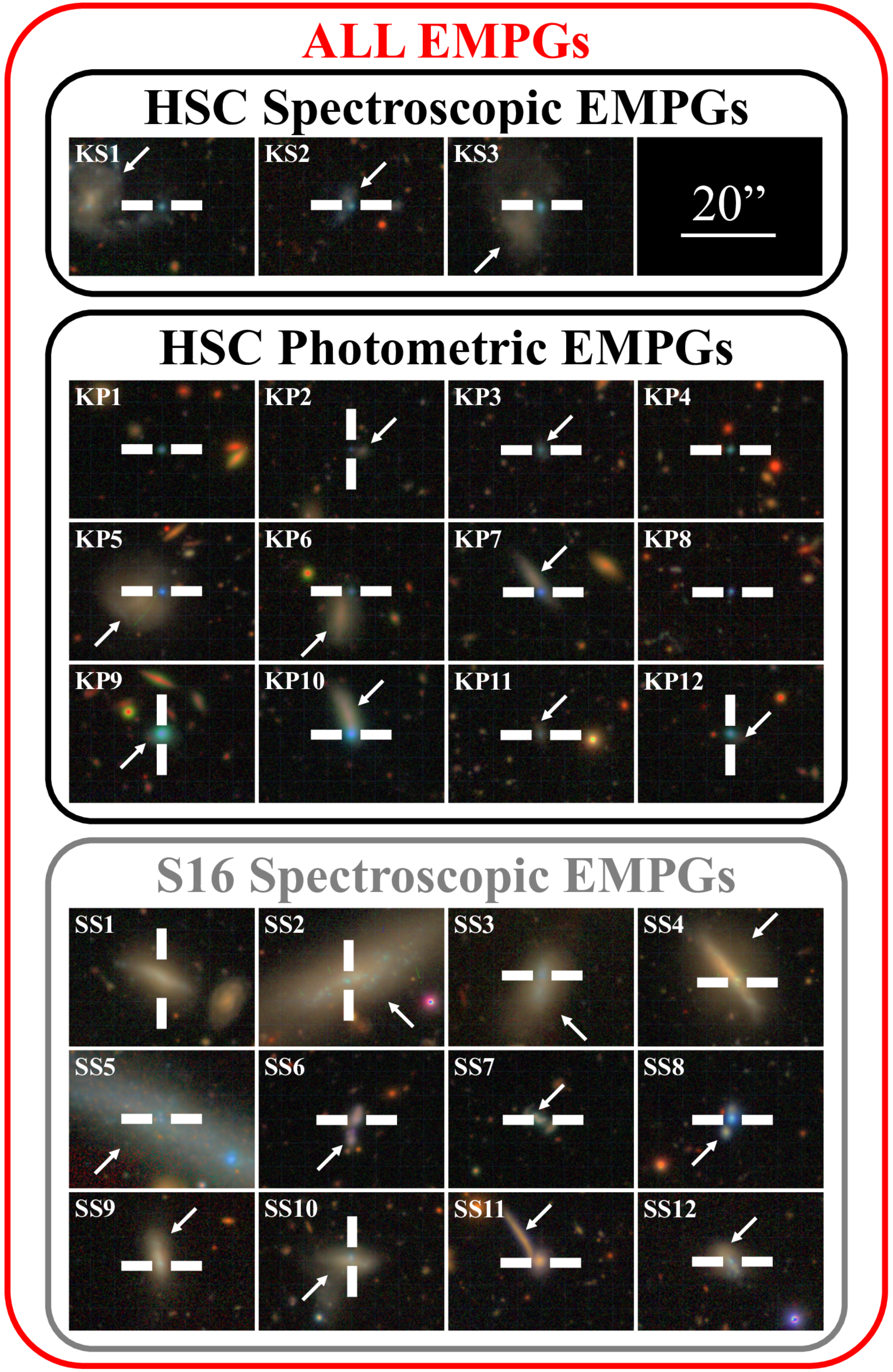}
    \caption{HSC $gri$-composite images of ALL EMPGs (Section \ref{subsec:empg}). \tcg{The HSC $g$, $r$, and $i$-bands correspond to blue, green, and red colors in the figure, respectively.} Each EMPG is located at the center of each panel. The cut-out size is $30"\times40"$. The arrows indicate the EMPG-tails. \tcg{Some of S16 spectroscopic EMPGs do not look blue because of the difference of redshifts and SFRs.}}
    \label{fig:morp}
\end{figure*}

In Sections \ref{subsec:hsc} we describe our imaging data. In Sections \ref{subsec:empg} and \ref{subsec:sa}, we explain our EMPG sample that we use for our study.

\subsection{HSC-SSP Imaging Data} \label{subsec:hsc}
We utilize the \tcr{imaging data set of HSC-SSP S18A} that were taken with 5 broadband filters, $grizy$, in 2014 March--2018 January\tcr{, and that were released by the HSC collaboration in 2018} \citep{Aihara2019}. In the HSC S18A imaging data, the effective area and the $i$-band limiting magnitude are $\sim500$ ${\rm deg}^2$ \tcg{(Paper I)} and $\sim26$ mag \tcg{(for point sources, \citealt{Ono2018})}, respectively. Because the HSC $y$-band image is $\sim1$ mag shallower than the other broadband images, we do not use the HSC $y$-band data but 4 broadband ($griz$) data for our analysis.

\subsection{\tcr{HSC} EMPGs} \label{subsec:empg}
\begin{figure*}[t]
    \centering
    \includegraphics[width=18.0cm]{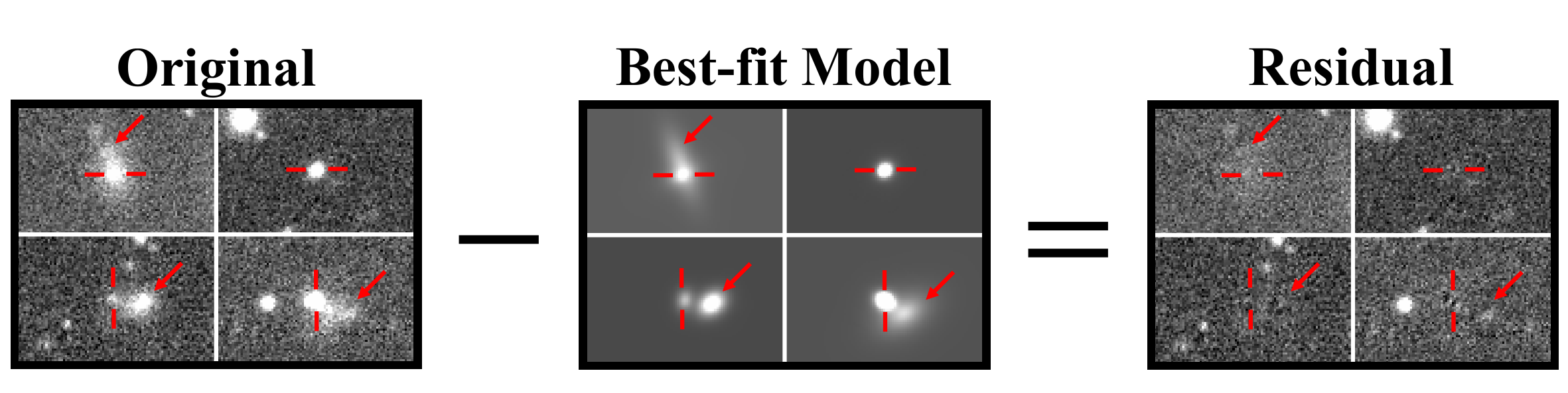}
    \caption{Examples of multi-component SB profile fitting. The left, middle, and right panels indicate the original images, the best-fit model images, and the residual ($={\rm original}-{\rm best}$-fit model) images, respectively.}
    \label{fig:galfit}
\end{figure*}

\tcr{Before we explain our sample, we need to clarify the photometric sample of Paper I. The Paper-I photometric sample consists of EMPG candidates identified with the data of HSC and SDSS, which are called HSC EMPG candidates and SDSS EMPG candidates, respectively. In this paper, we do not use SDSS EMPG candidates, because the SDSS EMPG candidates include \tcra{more contaminants} than the HSC EMPG candidates (Paper I). }
\tcr{The catalog of the HSC EMPG candidates} is developed with the HSC-SSP S17A and S18A data \citep{Aihara2019} that are wide and deep enough to search for rare and faint EMPGs. The HSC EMPG candidates are selected from $\sim46$ million sources whose photometric measurements are brighter than 5$\sigma$ limiting magnitudes in all of the 4 broadbands, $g<26.5$, $r<26.0$, $i<25.8$, and $z<25.2$ mag \tcg{\citep{Ono2018}}. The catalog consisting of these sources is referred to as the HSC source catalog.

With the HSC source catalog, Paper I isolates EMPGs from contaminants such as other types of galaxies, Galactic stars, and quasars. Paper I aims to find galaxies at $z\lesssim0.03$ with ${\rm EW}_{0}({\rm H}\alpha)>800$ ${\rm \AA}$ and $12+\log{(\rm O/H)}=6.69$--7.69. Because \tcg{it is difficult to} distinguish EMPGs from the contaminants on \tcg{2-color diagrams such as} $r-i$ vs. $g-r$, Paper I constructs a machine-learning classifier based on a deep neural network (DNN) with a training data set. The training data set is composed of mock photometric measurements for model spectra of EMPGs and the contaminants. The DNN allows us to isolate EMPGs from the contaminants with non-linear boundaries in the multi-dimensional color space. Paper I finally obtains 27 \tcr{HSC} EMPG candidates from the HSC source catalog. Paper I conducts spectroscopic follow-up observations for 4 out of the 27 \tcr{HSC} EMPG candidates, and confirm that all of the 4 \tcr{HSC} EMPG candidates are truly emission-line galaxies with the low metallicity of $12+\log{(\rm O/H)}=6.90$--8.27 (i.e., 1.6--38\% ${\rm Z}_{\odot}$). Because 2 out of the 4 \tcr{HSC} EMPG candidates meet the EMPG criterion of $12+\log{(\rm O/H)}<7.69$ (i.e., $<10\%$ ${\rm Z}_{\odot}$), Paper I concludes that these 2 candidates are quantitatively confirmed as EMPGs. There remain 2 ($=4-2$) spectroscopically confirmed \tcr{HSC} EMPG candidates with $12+\log{(\rm O/H)}=7.72$--8.27 (i.e., 11--38\% ${\rm Z}_{\odot}$). One of the 2 \tcr{HSC} EMPG candidates shows the low metallicity of $12+\log{(\rm O/H)}=7.72$ (i.e., $11\%$ ${\rm Z}_{\odot}$), almost meeting the EMPG criterion of $12+\log{(\rm O/H)}<7.69$ (i.e., $<10\%$ ${\rm Z}_{\odot}$). The other \tcr{HSC} EMPG candidate shows the moderately low metallicity of $12+\log{(\rm O/H)}=8.27$ (i.e., $38\%$ ${\rm Z}_{\odot}$), falling in the regime of metal-poor galaxies (MPGs). Thus the candidate is referred to as MPG, hereafter. 
There remain 23 ($=27-4$) \tcr{HSC} EMPG candidates that are not spectroscopically confirmed in Paper I. 
\tcr{
We obtain $r_{\rm e}$ for 15 of the 27 HSC EMPG candidates. Three out of the 15 HSC EMPG candidates have spectra, while 12 out of them do not have spectra.
}
We thus refer to the 3 EMPGs with $<11\%$ ${\rm Z}_{\odot}$ and 
the 12 EMPG candidates with no spectra as \tcr{HSC} spectroscopic EMPGs and \tcr{HSC} photometric EMPGs, respectively \tcg{(see Figure \ref{fig:morp})}. Hereafter we refer to the 15 objects as \tcr{HSC} EMPGs.

\tcr{Here we estimate the purity of the HSC photometric EMPGs. 
Paper I has reported only 4 out of the 27 HSC EMPG candidates that are spectroscopically observed, and the number of spectroscopic objects is too small to reliably estimate the purity. 
We have added other 13 HSC EMPG candidates whose spectra are recently taken with Keck/LRIS (Isobe et al. in prep.), and obtained 17 ($=4+13$) HSC EMPG candidates with spectroscopic results. We find that 12 out of the 17 HSC EMPG candidates with spectra ($12/17=71$\%) are real EMPGs with $\lesssim10$\% Z$_{\odot}$, while the rest of them ($5/17=29$\%) are contaminants. Thus, our machine learning classifier accomplishes 71\% purity for the HSC EMPG candidates, indicative that 71\% of the HSC photometric EMPGs are real EMPGs\footnote{Note that the HSC EMPGs include no contaminants on the basis of the spectroscopic observations.}. 
For the 12 spectroscopically-confirmed EMPGs, we derive the median redshift with the 16th and 84th percentiles $z=0.030_{-0.017}^{+0.015}$. \tcra{Thus, we assume $z=0.03$ for the HSC photometric EMPGs to measure their sizes, stellar masses, and SFRs.}}

\subsection{\tcg{S16 Spectroscopic EMPGs}} \label{subsec:sa}

\tcg{For the completeness of results, we also utilize a catalog of EMPGs made by S16. S16 select EMPGs in the full set of 788677 galaxy spectra at $z<0.25$ from the Sloan Digital Sky Survey (SDSS) Data Release 7. S16 aim to find galaxies showing a high line ratio of [{\sc O iii}]$\lambda$4363/[{\sc O iii}]$\lambda\lambda$4959, 5007, which is an indicator of gas with high electron temperature. Because metals are \tcr{the} main coolants of gas (e.g., \citealt{Pagel1979}), low-metallicity gas should \tcg{exhibit higher temperatures}. Using the automated classification algorithm, k-means (e.g., \citealt{Almeida2010}), S16 narrow down the large data set of the SDSS galaxy spectra according to the spectral shape in the range of $\lambda=4200$--5200 \AA, which contains [{\sc O iii}]$\lambda$4363, [{\sc O iii}]$\lambda$4959, and [{\sc O iii}]$\lambda$5007 lines. After the classification, S16 obtain 1281 EMPG candidates. S16 calculate metallicities of the 1281 candidates to identify that 196 out of the candidates meet the low metallicities of 2--10\% ${\rm Z}_{\odot}$, which meet the EMPG criterion. In order to evaluate $r_{\rm e}$ precisely in the same manner as HSC EMPGs, we utilize 13 out of the 196 EMPGs \tcg{that} have $griz$-band data \tcg{in} the HSC-SSP S18A data release. We obtain 12 EMPGs whose $r_{\rm e}$ values are successfully measured. Hereafter we refer to the 12 EMPGs as S16 spectroscopic EMPGs \tcg{(see Figure \ref{fig:morp})}.}

\subsection{\tcg{ALL EMPGs}}
\label{subsec:all}
\tcr{In order to increase the sample size,} we analyze both HSC EMPGs and S16 spectroscopic EMPGs.
We refer to the sum of HSC EMPGs and S16 spectroscopic EMPGs as \tcg{ALL EMPGs (see Figure \ref{fig:morp})}. Metallicities of ALL EMPGs\tcr{, so far identified,} are 0.01--0.1 Z$_{\odot}$.

\section{Analysis} \label{sec:analysis}
\begin{figure}[t]
    \centering
    \includegraphics[width=8.0cm]{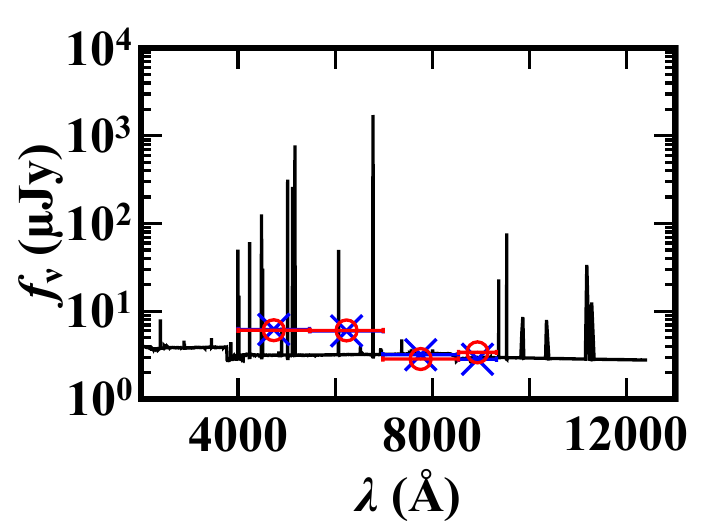}
    \caption{SED fitting result of \tcg{KS1}. The red circles, the solid black lines, and the blue crosses represent the observed photometric measurements, the best-fit spectrum, and the photometric measurements of the best-fit spectrum, respectively.}
    \label{fig:sed}
\end{figure}

In Section \ref{subsec:morp}, we report morphologies of ALL EMPGs. We present effective radii, stellar masses, and star formation rates (SFRs) of the EMPGs in Sections \ref{subsec:size}, \ref{subsec:mass}, and \ref{subsec:sfr}, respectively.

\subsection{Morphology} \label{subsec:morp}

Figure \ref{fig:morp} presents the HSC $gri$-composite images of ALL EMPGs. We find that most of the EMPGs have \tcr{tails}. \tcr{Hereafter we refer to the tails as EMPG-tails.} Here we define the \tcr{EMPG-tail} as \tcr{an object} brighter than the 5$\sigma$ limiting magnitude of the HSC $i$-band (25.8 mag) within 10 kpc from the EMPG. After conducting multi-component surface brightness (SB) profile fitting (Section \ref{subsec:size}), we find that \tcg{23 out of the 27 EMPGs} have \tcr{EMPG-tails, many of which appear to be galaxies in HSC deep images.} We \tcg{measure sizes and stellar masses of} the \tcr{EMPG-tails} in the same manner as the EMPGs.

\subsection{Size Measurement} \label{subsec:size}
\begin{figure}[t]
    \centering
    \includegraphics[width=8.0cm]{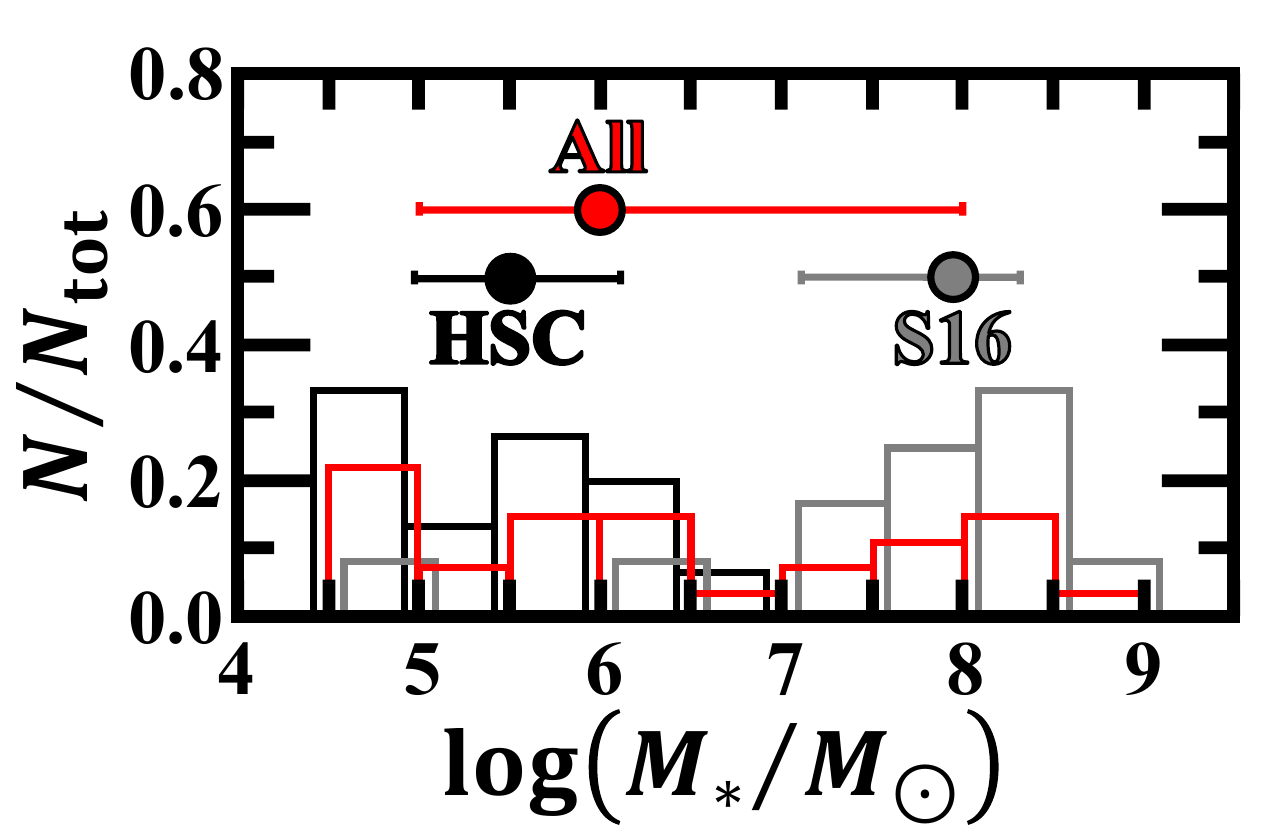}
    \caption{Normalized histogram of the stellar masses of the HSC EMPGs (black), S16 spectroscopic EMPGs (gray) \tcg{and ALL EMPGs (red)}. The number of EMPGs in each bin ($N$) is normalized by the total number of each sample ($N_{\rm tot}$). \tcg{For the presentation purpose, we slightly shift the histogram horizontally.} The black, gray, \tcg{and red filled circles with error bars} indicate median stellar masses of the HSC EMPGs (\tcra{$\log(M_{*}/{\rm M}_{\odot})=5.5$}), S16 spectroscopic EMPGs (\tcra{$\log(M_{*}/{\rm M}_{\odot})=7.9$}), \tcg{and ALL EMPGs (\tcra{$\log(M_{*}/{\rm M}_{\odot})=6.0$})}, respectively.}
    \label{fig:hist}
\end{figure}

We measure the galaxy size with the HSC $i$-band images. One of the reasons is that the $i$-band imaging data \tcg{allow us} to trace the spatial distribution of the stellar continuum because the $i$-band measurements are less affected by strong emission lines such as H$\alpha$ and {\sc [O iii]}. Another reason is that \tcg{the} median seeing of the HSC $i$-band images (${\rm FWHM}\sim0.56$ arcsec) is also smaller than \tcg{that} of the other four HSC broadband images \citep{Aihara2019}. We fit a S{\'e}rsic profile to the SB profile of each EMPG. The S{\'e}rsic profile can be written as
\begin{equation}
    \label{equ:sersic}
    I(r, n)=I_{\rm e}{\rm exp}\left[-\kappa(n)\left\{\left(\frac{r}{r_{\rm e}}\right)^{1/n}-1\right\}\right],
\end{equation}
where $r_{\rm e}$ and $n$ represent the effective radius and the S{\'e}rsic index, respectively. 
The function, $\kappa(n)$, is the implicit function that satisfies $\Gamma(2n)=2\gamma(2n,\kappa(n))$, where $\Gamma$ and $\gamma$ are the gamma function and the incomplete gamma function, respectively \citep{Graham2005}. The S{\'e}rsic profiles with $n=1$ and $n=4$ are generally obtained from the disk and elliptical galaxies, respectively. Because most of the EMPGs have EMPG-tails (Section \ref{subsec:morp}), we utilize the multi-component SB profile fitting code, {\sc Galfit} \citep{Peng2010}, to derive $r_{\rm e}$ and $n$ of the EMPGs (and the EMPG-tails simultaneously). 

Here we explain the procedure of our SB profile fitting. First, we fit the SB profiles of first single S{\'e}rsic profiles to those of the EMPGs by the $\chi^{2}$ minimization technique. \tcg{The code {\sc Galfit} can convolve the model functions with a point spread function (PSF), which is supplied by the HSC-SSP data release.} \tcg{We obtain best-fit models and residual images. The residual images are obtained by subtracting the best-fit model from an original image.} If there remain no obvious sources in the residual images, we complete the fitting of the EMPG with the best-fit model. If there exist clear sources in the residual images, we \tcg{execute two-component} S{\'e}rsic profile fitting\footnote{We include third single S{\'e}rsic profiles in our models if prominent sources in residual images affect fitting results.}. Physical properties of the first and the second S{\'e}rsic profiles are regarded as those of the EMPGs and EMPG-tails, respectively. The best-fit models provide physical properties of $r_{\rm e}$, $n$, \tcg{apparent $i$-band magnitudes} $m_{i}$, galaxy positions, axis ratios $q$, and position angles. We search $n$ in the range of $n=0.7$--4.2 because $n$ sometimes diverge at $n=0$ or $n\rightarrow\infty$. \tcg{We omit objects from the samples in cases where the SB profile is too complicated to fit.} We estimate the 16th, 50th, and 84th percentiles of the parameters by performing Monte-Carlo simulations. We create 100 mock images by cutting out each EMPG (and EMPG-tail) and embedding in nearby blank regions. We also consider that an error of each pixel is normally distributed with a variance value supplied by the HSC-SSP data release. Figure \ref{fig:galfit} presents examples of the SB profile fitting. The size and $n$ results are listed in Tables \ref{tab:empg}\tcg{-}\ref{tab:host}.

\subsection{Stellar Mass Estimation} \label{subsec:mass}
\begin{table*}[t]
    \centering
    \caption{Properties of the HSC EMPGs}
    \begin{tabular}{ccccccc} \hline
        Name & ID & Redshift & $r_{\rm e}$ & $n$ & $\log{(M_{*})}$ & $\log(\rm SFR)$ \\
        & & & pc & & ${\rm M}_{\odot}$ & ${\rm M}_{\odot}\:{\rm yr}^{-1}$ \\
        (1) & (2) & (3) & (4) & (5) & (6) & (7)\\ \hline \hline
        KS1 & J2314$+$0154 & 0.03265 & $104^{+12}_{-5}$ & $<0.7$ & \tcra{6.0} & $-0.851\pm0.001$ \\
        KS2 & J1631$+$4426 & 0.03125 & $137^{+9}_{-7}$ & $1.08^{+0.15}_{-0.13}$ & \tcra{6.1} & $-1.276\pm0.002$ \\
        KS3 & J1142$-$0038 & 0.02035 & $219^{+189}_{-13}$ & $2.79^{+1.41}_{-0.39}$ & \tcra{5.0} & $-1.066\pm0.002$ \\
        KP1 & J0912$-$0104 & --- & \tcr{$89.3^{+9.4}_{-12.4}$} & $0.83^{+0.37}_{-0.13}$ & \tcra{5.6} & \tcr{$-1.7^{+0.2}_{-0.4}$} \\
        KP2 & J2321$+$0125 & --- & \tcr{$377^{+87}_{-75}$} & $0.80^{+0.43}_{-0.10}$ & \tcra{4.8} & \tcr{$-2.0^{+0.2}_{-0.4}$} \\
        KP3 & J2355$+$0200 & --- & \tcr{$416^{+54}_{-51}$} & $0.80^{+0.14}_{-0.10}$ & \tcra{5.5} & \tcr{$-1.0^{+0.2}_{-0.4}$} \\
        KP4 & J2236$+$0444 & --- & \tcr{$159^{+13}_{-12}$} & $0.80^{+0.28}_{-0.10}$ & \tcra{5.9} & \tcr{$-1.7^{+0.2}_{-0.4}$} \\
        KP5 & J1411$-$0032 & --- & \tcr{$45.9^{+2.5}_{-1.7}$} & $0.7^{+0.4}_{-0.0}$ & \tcra{5.9} & \tcr{$-1.3^{+0.2}_{-0.4}$} \\
        KP6 & J0834$+$0103 & --- & \tcr{$116^{+8}_{-12}$} & $0.7^{+0.6}_{-0.0}$ & \tcra{4.9} & \tcr{$-1.6^{+0.2}_{-0.4}$} \\
        KP7 & J0226$-$0517 & --- & \tcr{$60.5^{+1.1}_{-11.7}$} & $1.16^{+0.91}_{-0.07}$ & \tcra{5.5} & \tcr{$-1.2^{+0.2}_{-0.4}$} \\
        KP8 & J0156$-$0421 & --- & \tcr{$199^{+23}_{-14}$} & $1.12^{+0.75}_{-0.42}$ & \tcra{5.0} & \tcr{$-1.6^{+0.2}_{-0.4}$} \\
        KP9 & J0935$-$0115 & --- & \tcr{$86.0^{+0.8}_{-1.3}$} & $<0.7$ & \tcra{6.8} & \tcr{$-0.4^{+0.2}_{-0.4}$} \\
        KP10 & J0937$-$0040 & --- & \tcr{$159\pm1$} & $1.72^{+0.03}_{-0.02}$ & \tcra{6.5} & \tcr{$-0.4^{+0.2}_{-0.4}$} \\
        KP11 & J1210$-$0103 & --- & \tcr{$589^{+227}_{-126}$} & $1.59^{+0.69}_{-0.38}$ & \tcra{5.0} & \tcr{$-1.6^{+0.2}_{-0.4}$} \\
        KP12 & J0845$+$0131 & --- & \tcr{$122^{+6}_{-7}$} & $1.72^{+2.48}_{-0.34}$ & \tcra{5.4} & \tcr{$-1.2^{+0.2}_{-0.4}$} \\ \hline
    \end{tabular}
    \tablecomments{(1): Name. (2): ID. (3): Spec-$z$. Typical uncertainties are $\Delta z\sim10^{-6}$ (Paper I). (4): Median effective radius with the 16th and 84th percentiles (Section \ref{subsec:size}). (5): Median S{\'e}rsic index with the 16th and 84th percentiles. \tcg{If we obtain $n=0.7$ for all of the mock images, we describe the situation as $n<0.7$} (Section \ref{subsec:size}). (6): Median stellar mass (Section \ref{subsec:mass}). (7): SFR (Section \ref{subsec:sfr}).}
    \label{tab:empg}
\end{table*}

We estimate stellar masses with the spectral energy distribution (SED) interpretation code, {\sc beagle} \citep{Chevallard2016}. The {\sc beagle} code calculates both the stellar continuum and the nebular emission using the stellar population synthesis code \citep{Bruzual2003} and the photoionization code of \citet{Gutkin2016} that are computed with {\sc cloudy} \citep{Ferland2013}. We adopt the \citet{Charlot2000} law to the models for dust attenuation. \tcg{In the SED fitting, we use griz-band photometry provided by the HSC-SSP S18A photometry catalog.} \tcg{Settings of the SED fitting for the EMPGs are the same as Paper I.} Because \tcg{Paper I} reports that most of our EMPGs with spectra show a \tcg{small} color excess of $E(B-V)\sim0$, we also assume no dust attenuation in the EMPGs. Assuming the constant star-formation history, we run the {\sc beagle} code with 4 free parameters of the metallicity, the maximum stellar age, the stellar mass, and the ionization parameter in the range of $Z=0.006$--0.3 ${\rm Z}_{\odot}$, $\log{({\rm Age}/{\rm yr})}=4.0$--\tcra{9.0}, $\log{(M_{*}/{\rm M}_{\odot})}=4.0$--9.0, and $\log{(U)}=(-2.5)$--$(-0.5)$, respectively. 
\tcr{This time we assume the maximum stellar age of the EMPGs less than \tcra{1 Gyr}, because the EMPGs do not show prominent Balmer breaks (Paper I) indicative that the EMPGs are much younger than $\sim1$ Gyr.} 
An example of the SED fitting is shown in Figure \ref{fig:sed}. \tcg{We also conduct SED fitting for the EMPG-tails, while parameter ranges of the fitting are different from those for the EMPGs.} Assuming the constant star-formation history, we run the {\sc beagle} code with 5 free parameters of the metallicity, the maximum stellar age, the stellar mass, the ionization parameter, and the dust attenuation in the range of $Z=0.01$--1 ${\rm Z}_{\odot}$, $\log{({\rm Age}/{\rm yr})}=8.0$--12.0, $\log{(M_{*}/{\rm M}_{\odot})}=4.0$--9.0, $\log{(U)}=(-5.0)$--$(-2.5)$, and $\tau_{\rm V}=0$--20, respectively. \tcg{We find that 8 EMPG-tails are missed in the HSC-SSP photometry catalog \tcg{probably because the 8 EMPG-tails are not only faint ($\sim20$ mag) but also near their EMPG ($\sim1$ arcsec)}, while the other \tcg{15 ($=23-8$)} EMPG-tails are included. We first estimate stellar masses of the 15 EMPG-tails by the SED fitting described above.
Then we obtain a mass-luminosity ($M_{*}$ and absolute $i$-band luminosity $M_{i}$) relation, 
\begin{equation}
    \label{equ:ml}
    \log(M_{*}/{\rm M}_{\odot})=\tcr{-0.345M_{i}+2.02},
\end{equation}
by the linear fitting to the stellar masses and $i$-band luminosities of the 15 EMPG-tails. 
For the 8 EMPG-tails missed in the HSC-SSP photometry catalog, we instead use $i$-band magnitudes obtained by our SB profile fitting (Section \ref{subsec:size}). 
Then, we apply the mass-luminosity relation (Equation \ref{equ:ml}) to estimate stellar masses of the 8 EMPG-tails.} 

\tcra{Stellar masses of the EMPGs and the EMPG-tails are gathered in Tables \ref{tab:empg}-\ref{tab:host}. In these tables, we show only median values of the stellar masses because errors provided by the SED fitting do not include any uncertainty arising from different assumptions. This uncertainty is $\sim0.1$ dex, which is larger than a typical error of $\sim0.05$ dex provided by the SED fitting. }

\subsection{SFR} \label{subsec:sfr}

\begin{table*}[t]
    \centering
    \caption{Properties of S16 spectroscopic EMPGs}
    \begin{tabular}{cccccccc} \hline
        Name & \# in & ID & Redshift & $r_{\rm e}$ & $n$ & $\log{(M_{*})}$ & $\log(\rm SFR)$ \\
        & S16 & & & pc & & ${\rm M}_{\odot}$ & ${\rm M}_{\odot}\:{\rm yr}^{-1}$ \\
        (1) & (2) & (3) & (4) & (5) & (6) & (7) & (8)\\ \hline \hline
        SS1 & 110 & J1217$-$0154 & 0.02047 & $1520^{+30}_{-20}$ & $1.29\pm0.03$ & \tcra{8.0} & $-1.959\pm0.004$ \\
        SS2 & 148 & J1419$+$0109 & 0.00814 & $80.8^{+0.5}_{-1.9}$ & $<0.7$ & \tcra{7.2} & $-2.210\pm0.005$ \\
        SS3 & 152 & J1427$-$0143 & 0.00602 & $343^{+10}_{-11}$ & $1.38^{+0.03}_{-0.04}$ & \tcra{6.0} & $-2.660\pm0.004$ \\
        SS4 & 153 & J1429$+$0107 & 0.02969 & $422^{+1}_{-0}$ & $0.92\pm0.00$ & \tcra{8.3} & $-0.975\pm0.003$ \\
        SS5 & 158 & J1444$+$4237 & 0.00213 & $37.7^{+8.1}_{-6.4}$ & $4.2^{+0.0}_{-0.5}$ & \tcra{5.0} & $-3.560\pm0.003$ \\
        SS6 & 186 & J2211$+$0048 & 0.06459 & $662^{+4}_{-2}$ & $0.85\pm0.02$ & \tcra{8.3} & $-0.883\pm0.003$ \\
        SS7 & 187 & J2212$+$0108 & 0.21011 & $2650\pm30$ & $1.50^{+0.04}_{-0.03}$ & \tcra{8.5} & $0.418\pm0.005$ \\
        SS8 & 191 & J2302$+$0049 & 0.03312 & $273\pm1$ & $1.38^{+0.05}_{-0.02}$ & \tcra{7.4} & $-0.631\pm0.004$ \\
        SS9 & 192 & J2327$-$0051 & 0.02343 & $494^{+3}_{-2}$ & $1.01\pm0.01$ & \tcra{8.0} & $-1.730\pm0.005$ \\
        SS10 & 193 & J2334$+$0029 & 0.02384 & $684^{+64}_{-88}$ & $>4.2$ & \tcra{7.9} & $-1.461\pm0.004$ \\
        SS11 & 194 & J2335$-$0025 & 0.07672 & $1260^{+10}_{-0}$ & $1.18^{+0.02}_{-0.01}$ & \tcra{8.6} & $-0.595\pm0.005$ \\
        SS12 & 195 & J2340$-$0053 & 0.01883 & $176^{+2}_{-4}$ & $<0.7$ & \tcra{7.7} & $-1.747\pm0.005$ \\ \hline
    \end{tabular}
    \tablecomments{(1): Name. (2): Number appeared in S16. (3): ID. (4): Spec-$z$. Typical uncertainties are $\Delta z\lesssim10^{-5}$. (5): Median effective radius with the 16th and 84th percentiles (Section \ref{subsec:size}). (6): Median S{\'e}rsic index with the 16th and 84th percentiles (Section \ref{subsec:size}). (7): Median stellar mass (Section \ref{subsec:mass}). (8): SFR (Section \ref{subsec:sfr}).}
    \label{tab:sa}
\end{table*}

\begin{table}[t]
    \centering
    \caption{Coordinates of the HSC spectroscopic EMPGs and S16 spectroscopic EMPGs}
    \begin{tabular}{cccc} \hline
        Name & ID & R.A. & Dec. \\
        & & (hh:mm:ss) & (dd:mm:ss) \\
        (1) & (2) & (3) & (4) \\ \hline \hline
        KS1 & J2314$+$0154 & 23:14:37.6 & $+01$:54:14.3 \\
        KS2 & J1631$+$4426 & 16:31:14.2 & $+44$:26:04.4 \\
        KS3 & J1142$-$0038 & 11:42:25.2 & $-00$:38:55.6 \\
        SS1 & J1217$-$0154 & 12:17:10.2 & $-01$:54:25.6 \\
        SS2 & J1419$+$0109 & 14:19:20.2 & $+01$:09:54.9 \\
        SS3 & J1427$-$0143 & 14:27:04.8 & $-01$:43:46.9 \\
        SS4 & J1429$+$0107 & 14:29:32.6 & $+01$:07:02.2 \\
        SS5 & J1444$+$4237 & 14:44:12.8 & $+42$:37:44.0 \\
        SS6 & J2211$+$0048 & 22:11:17.9 & $+00$:48:05.0 \\
        SS7 & J2212$+$0108 & 22:12:26.9 & $+01$:08:35.3 \\
        SS8 & J2302$+$0049 & 23:02:10.0 & $+00$:49:38.8 \\
        SS9 & J2327$-$0051 & 23:27:30.5 & $-00$:51:14.6 \\
        SS10 & J2334$+$0029 & 23:34:14.8 & $+00$:29:07.3 \\
        SS11 & J2335$-$0025 & 23:35:40.7 & $-00$:25:33.1 \\
        SS12 & J2340$-$0053 & 23:40:38.4 & $-00$:53:30.8 \\
        \hline
    \end{tabular}
    \tablecomments{(1): Name. (2): ID. (3): R.A. in J2000. (4): Dec. in J2000.}
    \label{tab:coord}
\end{table}

We derive SFRs from H$\alpha$ fluxes $F({\rm H}\alpha)$. Regarding the HSC spectroscopic EMPGs, we \tcg{use dust- \tcr{and aperture-}corrected H$\alpha$ fluxes obtained by spectroscopy in Paper I. For S16 spectroscopic EMPGs, we utilize \tcr{aperture-corrected H$\alpha$ fluxes of SDSS DR16 that are derived with the methods of \citet{Brinchmann2004}, \citet{Kauffmann2003}, and \citet{Tremonti2004}.} On the other hand, we estimate H$\alpha$ fluxes of the HSC photometric EMPGs with $riz$-band photometry.} The HSC EMPGs show large $r$-band excesses by $r-i\sim-0.8$, which are mainly caused by the strong H$\alpha$ line (Section \ref{subsec:empg}). 
Here we describe how to estimate the flux density (per unit \tcr{frequency}) of the $r$-band continuum $f_{r, {\rm cont}}$. 
Because the observed $iz$-band \tcr{flux densities}, $f_{i, {\rm tot}}$ and $f_{z, {\rm tot}}$, are less affected by strong emission lines (Section \ref{subsec:size}), we regard $f_{i, {\rm tot}}$ and $f_{z, {\rm tot}}$ as tracers of the stellar continuum.
\tcr{\citet{Bruzual1993} synthesis model shows that young starburst galaxies have stellar continua whose flux densities per unit frequency is constant. We thus estimate $f_{r, {\rm cont}}$ as an average of $f_{i, {\rm tot}}$ and $f_{z, {\rm tot}}$. We calculate $F({\rm H}\alpha)$ as follows:}
\begin{equation}
    \label{equ:fha}
    F({\rm H}\alpha)=(f_{r, {\rm tot}}-f_{r, {\rm cont.}})\tcr{\times\frac{c}{\lambda_{r}^{2}}\times}\Delta\lambda_{r},
\end{equation}
where $f_{r, {\rm tot}}$, \tcr{$\lambda_{r}$} and $\Delta\lambda_{r}$ represent the observed $r$-band \tcr{flux density, the central wavelength of the HSC $r$-band filter,} and \tcg{the} width of the HSC $r$-band filter in wavelength, respectively. \tcr{The $F({\rm H}\alpha)$ error calculated from $riz$-band photometric errors is at most $\sim10$\%. Emission lines other than H$\alpha$ in the HSC $r$, $i$, and $z$ bands can increase the flux densities by only $\sim11$, 4, and 6\%, respectively. Consequently, the estimated H$\alpha$ flux can be changed by only $\sim6$\%.} We calculate \tcg{$F({\rm H}\alpha)$} of the HSC spectroscopic EMPGs, and confirm that the values of \tcg{$F({\rm H}\alpha)$} are consistent with those derived from the spectra \tcr{within $\sim60$\%. Because this $\sim60$\% difference is the most dominant error of the $F({\rm H}\alpha)$ estimation, we adopt $\pm60$\% as the $F({\rm H}\alpha)$ error of the HSC photometric EMPGs.} We utilize the Kennicutt relation \citep{Kennicutt1998} to derive SFRs:
\begin{equation}
    \label{equ:Ken98}
    {\rm SFR}=7.9\times10^{-42}L({\rm H}\alpha),
\end{equation}
where SFR and $L({\rm H}\alpha)$ are in units of ${\rm M}_{\odot}\:{\rm yr}^{-1}$ and erg ${\rm s}^{-1}$, respectively. We note that \citet{Kennicutt1998} adopt the power-law initial mass function (IMF) of \citet{Salpeter1955} to derive Equation \ref{equ:Ken98}. However, the top-heavy \citet{Chabrier2003} IMF is more appropriate than the Salpeter IMF for young galaxies such as the HSC EMPGs (Paper I). We divide SFR of Equation \ref{equ:Ken98} by 1.8, which is based on the Chabrier IMF \citep{Madau2014}. The SFR results are listed in Tables \ref{tab:empg} and \ref{tab:sa}.

\section{Results} \label{sec:results}
In Section \ref{subsec:prop}, we report the \tcg{morphological} properties of the EMPGs and EMPG-tails. In Sections \ref{subsec:remstar} and \ref{subsec:sfrmstar}, we describe the relations among the properties. 
In Section \ref{subsec:remstar}, we report the $r_{\rm e}$--$M_{*}$ relation to compare the EMPGs and EMPG-tails to local galaxies and high-$z$ low-mass galaxies. In Section \ref{subsec:sfrmstar}, we present the SFR--$M_{*}$ relation to show the star-forming activities of the EMPGs.  

\subsection{\tcg{Size, S{\'e}rsic Index, Stellar Mass, and SFR}} \label{subsec:prop}

\begin{table}[t]
    \centering
    \caption{Properties of the EMPG-tails}
    \begin{tabular}{cccc} \hline
        Name & $r_{\rm e}$ & $n$ & $\log{(M_{*})}$ \\
        & kpc & & ${\rm M}_{\odot}$ \\
        (1) & (2) & (3) & (4)\\ \hline \hline
        KS1-tail & $3.37^{+0.34}_{-0.29}$ & $2.51^{+0.18}_{-0.15}$ & 7.5 \\
        KS2-tail & $0.805^{+0.063}_{-0.054}$ & $<0.7$ & 6.9 \\
        KS3-tail & $3.99^{+1.67}_{-0.64}$ & $1.93^{+0.37}_{-0.16}$ & 8.0 \\
        KP2-tail & \tcr{$0.481^{+0.023}_{-0.016}$} & $<0.7$ & 6.5 \\
        KP3-tail & \tcr{$0.705^{+0.104}_{-0.045}$} & $<0.7$ & 6.3 \\
        KP5-tail & \tcr{$2.62\pm0.05$} & $0.93\pm0.02$ & 8.0 \\
        KP6-tail & \tcr{$1.98^{+0.07}_{-0.02}$} & $0.97^{+0.05}_{-0.02}$ & 8.4 \\
        KP7-tail & \tcr{$1.87^{+0.01}_{-0.02}$} & $0.79^{+0.04}_{-0.03}$ & 7.9 \\
        KP9-tail & \tcr{$1.23^{+0.11}_{-0.07}$} & $0.95^{+0.12}_{-0.08}$ & 7.1 \\
        KP10-tail & \tcr{$1.63^{+0.01}_{-0.00}$} & $<0.7$ & 8.1 \\
        KP11-tail & \tcr{$1.29^{+0.18}_{-0.11}$} & $0.7^{+0.1}_{-0.0}$ & 6.5 \\
        KP12-tail & \tcr{$1.02^{+0.45}_{-0.11}$} & $0.89^{+0.34}_{-0.19}$ & 6.6 \\
        SS2-tail & $6.66^{+0.09}_{-0.53}$ & $2.10^{+0.03}_{-0.07}$ & 7.3 \\
        SS3-tail & $0.822^{+0.026}_{-0.028}$ & $1.55^{+0.04}_{-0.03}$ & 6.2 \\
        SS4-tail & $2.77^{+0.02}_{-0.03}$ & $1.60\pm0.02$ & 8.8 \\
        SS5-tail & $1.84^{+0.11}_{-0.21}$ & $<0.7$ & 7.4 \\
        SS6-tail & $1.64^{+0.03}_{-0.02}$ & $1.43^{+0.03}_{-0.02}$ & 7.8 \\
        SS7-tail & $1.90^{+0.26}_{-0.18}$ & $4.2^{+0.0}_{-0.2}$ & 8.3 \\
        SS8-tail & $0.321\pm0.002$ & $0.89^{+0.03}_{-0.01}$ & 7.1 \\
        SS9-tail & $1.31^{+0.03}_{-0.01}$ & $1.22^{+0.03}_{-0.02}$ & 7.9 \\
        SS10-tail & $1.41^{+0.04}_{-0.00}$ & $1.02^{+0.04}_{-0.01}$ & 7.9 \\
        SS11-tail & $4.23\pm0.02$ & $<0.7$ & 8.8 \\
        SS12-tail & $0.645^{+0.008}_{-0.007}$ & $0.77\pm0.01$ & 7.5 \\ \hline
    \end{tabular}
    \tablecomments{(1): Name. (2): Median effective radius with the 16th and 84th percentiles (Section \ref{subsec:size}). (3): Median S{\'e}rsic index with the 16th and 84th percentiles (Section \ref{subsec:size}). (4): Median stellar mass (Section \ref{subsec:mass}).}
    \label{tab:host}
\end{table}

\tcg{Regarding ALL EMPGs, we obtain a median effective radius of \tcr{$r_{\rm e}=200^{+450}_{-110}$} pc, S{\'e}rsic index of $n=1.1^{+0.6}_{-0.4}$, stellar mass of \tcra{$\log(M_{*}/{\rm M}_{\odot})=6.0^{+2.0}_{-1.0}$}, and SFR of \tcr{$\log({\rm SFR}/{\rm M}_{\odot}\:{\rm yr}^{-1})=-1.3^{+0.7}_{-0.6}$} with the range of $\pm68$\% distributions, respectively.} The small values of \tcg{$r_{\rm e}\sim200$ pc} and $n\sim1$ suggest that the EMPGs \tcg{have} very compact disks. The \tcg{median} size of the EMPGs is also comparable to
\tcr{those of the heads of metal-poor tadpole galaxies (${\rm FWHM}\sim200$ pc; \citealt{SanchezAlmeida2015}) even though their method of measuring and the interpretation of the morphological structure are different from ours}. As shown in Figure \ref{fig:hist}, the median $M_{*}$ of the HSC EMPGs is $\sim2$ dex smaller than that of S16 spectroscopic EMPGs, which makes the EMPGs cover the wide $M_{*}$ range of $\log(M_{*}/{\rm M}_{\odot})=4.5$--\tcra{8.6}.
The HSC EMPGs have small $M_{*}$ comparable to those of Galactic star clusters. All the results are listed in Tables \ref{tab:empg} and \ref{tab:sa}.

On the other hand, we \tcg{find} that a median effective radius, S{\'e}rsic index, and stellar mass of the EMPG-tails with the range of $\pm68$\% distributions are \tcr{$r_{\rm e}=1.6^{+1.4}_{-0.9}$} kpc, 
$n=0.9^{+0.8}_{-0.2}$, and \tcr{$\log(M_{*}/{\rm M}_{\odot})=7.5^{+0.7}_{-0.9}$}, respectively (Table \ref{tab:host}).
\tcra{The stellar mass ratio between EMPGs and EMPG-tails
($M_{\rm *, tail}/M_{\rm *, EMPG}\sim32$) is comparable to the ratio between head and tail of tadpole galaxies ($M_{\rm *, tail}/M_{\rm *, head}\sim12$; \citealt{Elmegreen2012}), which implies that heads of the tadpole galaxies and the EMPG with the EMPG-tail are similar populations.}

\subsection{Size-Stellar Mass Relation} \label{subsec:remstar}
\begin{figure*}[t]
    \centering
    \includegraphics[width=18.0cm]{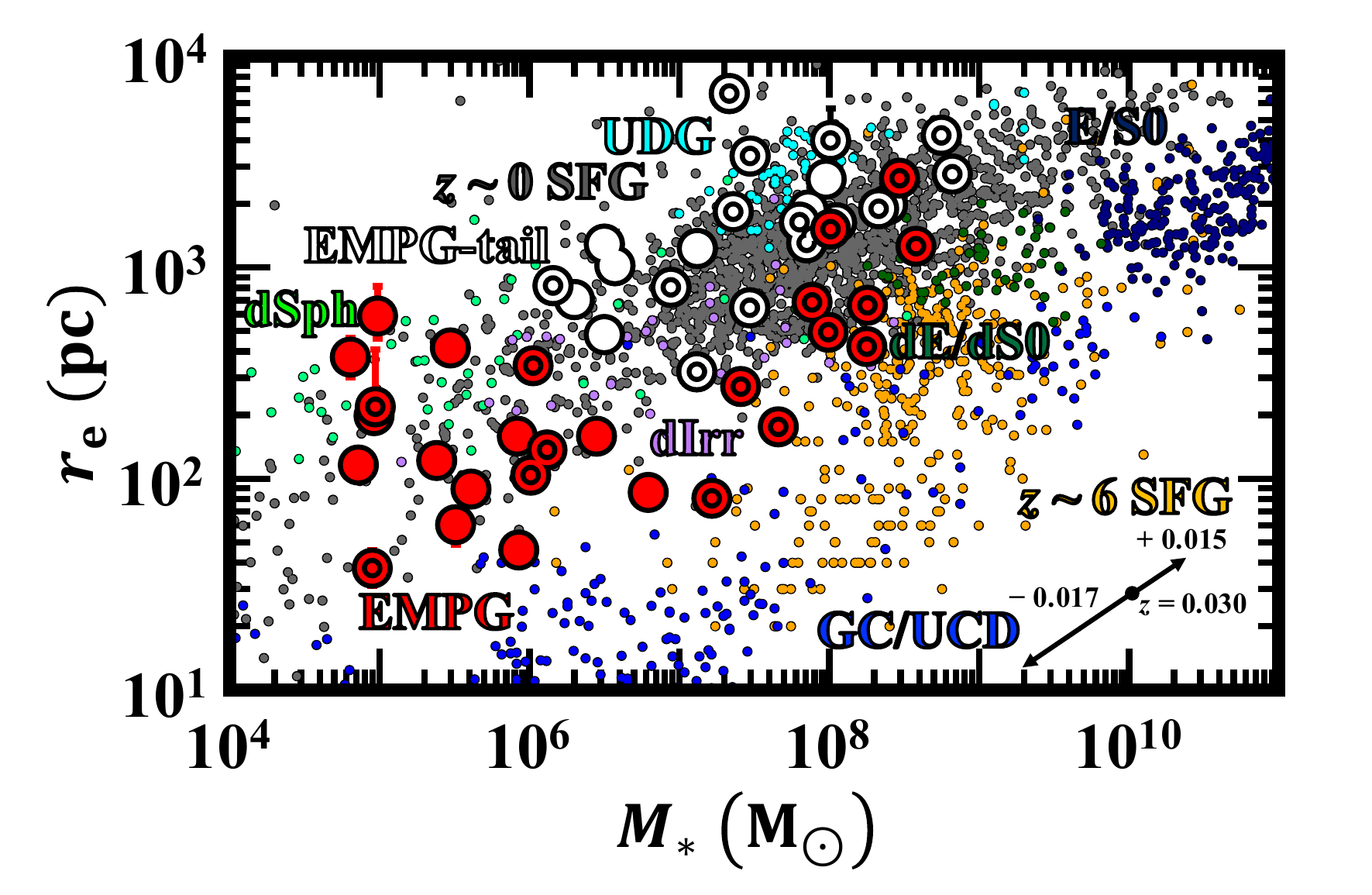}
    \caption{Distribution of $r_{\rm e}$ and $M_{*}$ of the EMPGs (red) and the EMPG-tails (white). The EMPGs with spectra (and their EMPG-tail) are represented as the double-edged circle. We add the data of SFGs at $z\sim0$ (gray; \citealt{Shibuya2015}) and $z\sim6$ (yellow; \citealt{Shibuya2015}; \citealt{kikuchihara2020}). We also plot $r_{\rm e}$ and $M_{*}$ of local dwarf galaxies such as ellipticals (E/S0; navy; \citealt{Norris2014}), dwarf ellipticals (dE/dS0; dark green; \citealt{Norris2014}), globular clusters or ultra compact dwarfs (GC/UCD; blue; \citealt{Norris2014}), dwarf irregulars (dIrr; purple; \citealt{McConnachie2012}), dwarf spheroidals (dSph; light green; \citealt{McConnachie2012}), and UDGs (cyan; \citealt{VanDokkum2015}; \citealt{Hashimoto2020}). The two arrows at the right corner indicate how the plots of the HSC EMPGs can change when the assumed $z$ varies in the range of \tcr{$\Delta z=[-0.017,+0.015]$}.}
    \label{fig:remstar}
\end{figure*}

\begin{figure*}[t]
    \centering
    \includegraphics[width=18.0cm]{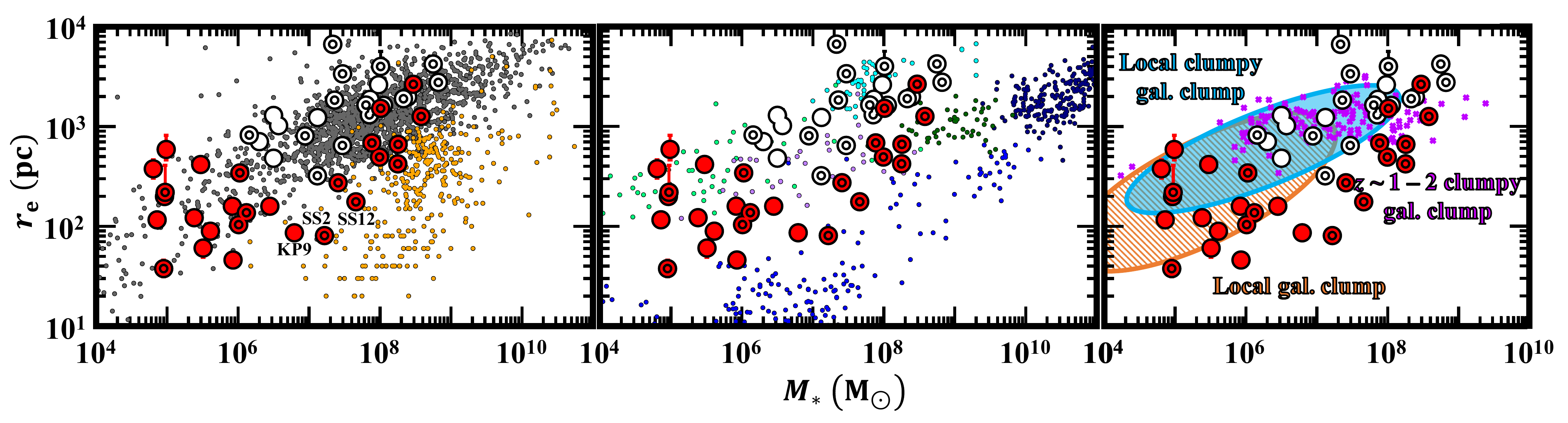}
    \caption{(Left) Same as Figure \ref{fig:remstar}, but we limit to $z\sim0$ and 6 SFGs. (\tcr{Center}) Same as Figure \ref{fig:remstar}, but we limit to local dwarf galaxies. \tcr{(Right) The EMPGs and EMPG-tails with clumps of $z\sim1$--2 clumpy galaxies (magenta cross), local clumpy galaxies (light-blue oval) and normal galaxies (orange striped oval). The data of the all clumps are taken from \citet{Elmegreen2013}.}}
    \label{fig:remstar_feat}
\end{figure*}

Figure \ref{fig:remstar} represents the distribution of $r_{\rm e}$ and $M_{*}$ of the EMPGs and EMPG-tails. We add the data of star-forming galaxies (SFGs) at $z\sim0$ (gray) and $z\sim6$ (yellow). We also plot $r_{\rm e}$ and $M_{*}$ of local dwarf galaxies. We make Figure \ref{fig:remstar_feat} to compare the distributions of the EMPGs and EMPG-tails to the $z\sim0$ and 6 SFGs (left), the local dwarf galaxies (\tcr{center})\tcr{, and clumps of clumpy galaxies and normal galaxies (right)}. As described in the left panel of Figure \ref{fig:remstar_feat}, we \tcg{find} that most of the EMPGs, except for a few (\tcra{KP9, SS2, and SS12}), fall on the $r_{\rm e}$--$M_{*}$ relation of $z\sim0$ SFGs rather than $z\sim6$ SFGs. The EMPGs have the $r_{\rm e}$ values larger than those of $z\sim6$ SFGs at a given $M_{*}$. 
Compared to local dwarf galaxies as shown in the \tcr{center} panel of Figure \ref{fig:remstar_feat}, some of the EMPGs \tcg{have the values of $r_{\rm e}$ and $M_{*}$ similar to} those of dSphs and dIrrs. The other EMPGs fall on the region between dSphs and GCs. 
\tcr{Comparing the right panel of Figure \ref{fig:remstar_feat} with the left panel of Figure \ref{fig:remstar_feat}, we find that the clumps of \citet{Elmegreen2013} have size--mass relations similar to those of $z\sim0$ SFGs. Thus we can say the same things as we compare the EMPGs with $z\sim0$ SFGs. It should be noted that the sizes of the clumps reported by \citet{Elmegreen2013} do not necessarily represent effective radii. }

As well as the EMPGs, the majority of the EMPG-tails have the $r_{\rm e}$--$M_{*}$ relation similar to that of $z\sim0$ SFGs. We also find that the EMPG-tails are located \tcg{on} the distributions of dSphs, dIrrs, and UDGs.

\subsection{SFR-Stellar Mass Relation} \label{subsec:sfrmstar}
\begin{figure}[t]
    \centering
    \includegraphics[width=8.0cm]{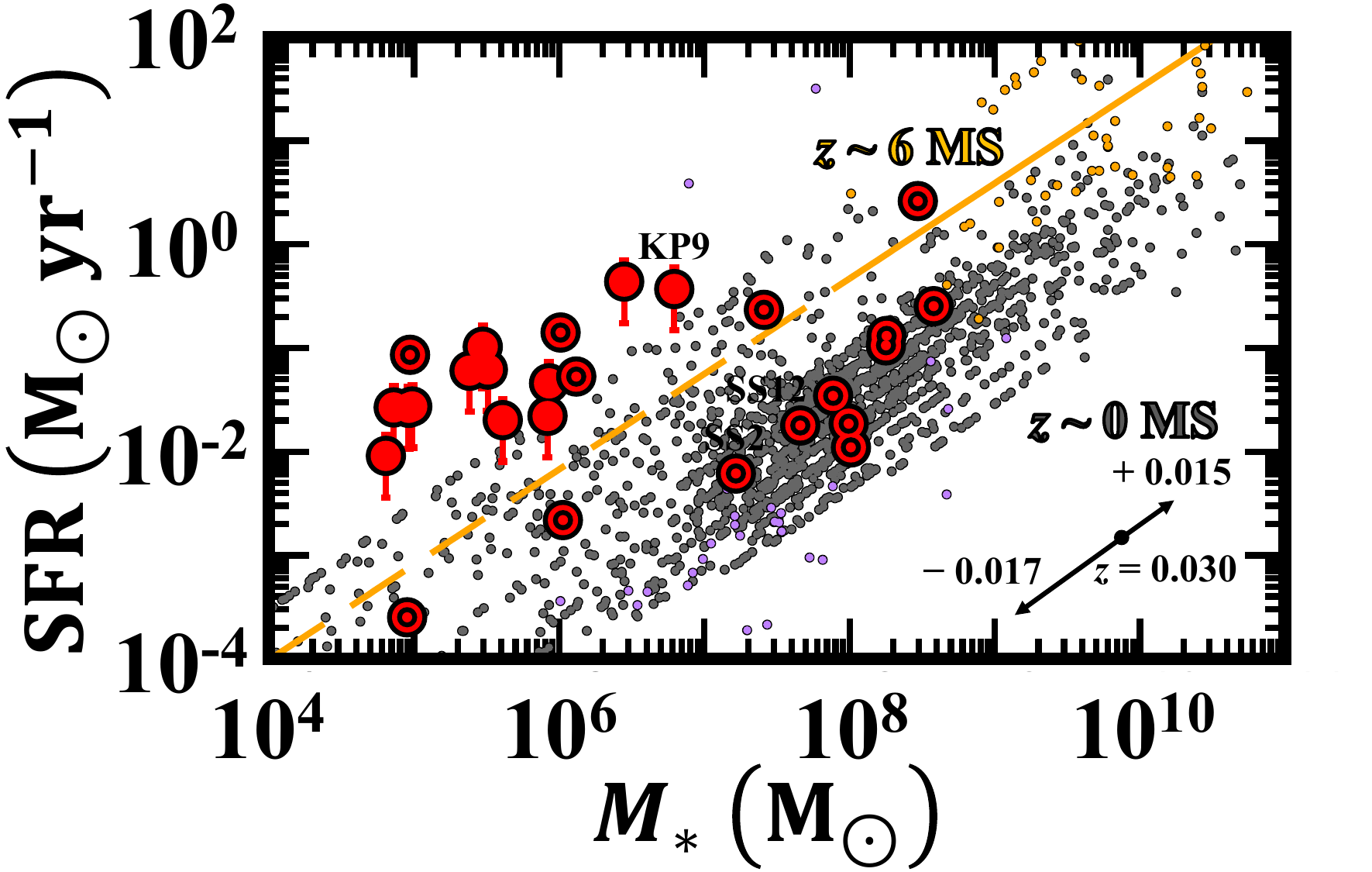}
    \caption{Distribution of SFR and $M_{*}$ of the EMPGs. The symbols are the same as in Figure \ref{fig:remstar}. \tcg{The distribution of the gray filled circles describe the low-mass regime of the $z\sim0$ star-formation main sequence (MS). We add a $5<z<6$ MS of \citet{Santini2017} represented as the yellow solid line. We extrapolate the MS of \citet{Santini2017} to the low-mass regime of $\log(M_{*}/M_{\odot})<10^{8}$, which is shown as the yellow dashed line.}}
    \label{fig:sfrmstar}
\end{figure}

In Section \ref{subsec:remstar}, we report that the EMPGs overlap the distributions of $z\sim0$ SFGs on the $r_{\rm e}$--$M_{*}$ plane (Figure \ref{fig:remstar_feat} (Left)). Now we present the SFR--$M_{*}$ distribution of the EMPGs to compare with those of $z\sim0$ and 6 main sequences (MSs) \citep{Shibuya2015}. As shown in Figure \ref{fig:sfrmstar}, the EMPGs fall on both the $z\sim0$ MS and the extrapolation of $z\sim6$ MS. 
\tcra{We confirm that KP9 is located around the extrapolation of the $z\sim6$ MS, whereas SS2 and SS12 lie on the $z\sim0$ MS.}

\tcg{In Section \ref{subsec:remstar}, we also point out that some of the EMPGs have $r_{\rm e}$ and $M_{*}$ comparable to those of dSphs and dIrrs. Now we compare the EMPGs to dIrrs whose SFR and $M_{*}$ values are reported by \citet{Zhang2012}. As shown in Figure \ref{fig:sfrmstar}, the majority of the EMPGs, except for a few, have sSFR values higher than those of dIrrs. We note that star formation activities of dSphs are already quenched \citep{Weisz2014}, i.e., SFR values of dSphs are too small to plot.}

\section{Discussions} \label{sec:discuss}
\subsection{EMPG} \label{subsec:discuss_empg}
In Section \ref{subsec:remstar}, we compare the EMPGs to several types of galaxies \tcr{and clumps} \tcg{in the $r_{\rm e}$--$M_{*}$ space}. In this section, we discuss which type of galaxies can be a counterpart of the EMPG based on the properties that we report in Section \ref{sec:results}. 

\subsubsection{Comparison with SFGs} \label{subsubsec:sfg}
In Section \ref{sec:intro}, we introduce the idea that EMPGs are expected to be local analogs of high-$z$ young galaxies. 
However, in Section \ref{subsec:remstar} we report that most of the EMPGs have the $r_{\rm e}$ values larger than those of $z\sim6$ SFGs for a given $M_{*}$, which suggests that not every EMPG is a perfect local analog of high-$z$ young galaxies. 
KP9 is only an exception whose $r_{\rm e}$, $M_{*}$ and SFR are similar to those of $z\sim6$ SFGs (Sections \ref{subsec:remstar} and \ref{subsec:sfrmstar}), which suggests that KP9 can be a local analog of high-$z$ young galaxies. 
\tcg{However, it might be natural that high-$z$ young galaxies are generally more compact than local galaxies. Considering that the slope of $r_{\rm e}$--$M_{*}$ relation of SFGs do not significantly evolve toward high-$z$, \citet{VanDerWel2014} conclude that the sizes of SFGs are determined by \tcg{the sizes of} the host dark-matter (DM) halos. This result suggests that high-$z$ SFGs should be more compact than local SFGs when other parameters \tcg{(e.g., $M_{*}$)} are the same. Thus, $r_{\rm e}$ values of the EMPGs might inevitably be larger than those of $z\sim6$ SFGs if the EMPGs are local analogs of high-$z$ young galaxies in reality. However, there is no evidence that we can adopt the trend that $r_{\rm e}$ values decrease toward high redshifts for galaxies in the low-mass regime of $\log(M_{*}/{\rm M}_{\odot})\lesssim7$. This problem will be solved by either surveys for high-$z$ low-mass galaxies or high-resolution cosmological zoom-in simulations of low-mass galaxies.}

On the other hand, we find that most of the EMPGs have the $r_{\rm e}$--$M_{*}$ relation similar to those of $z\sim0$ SFGs (Section \ref{subsec:remstar}). \tcg{We also \tcg{find} that some of the EMPGs fall on the $z\sim0$ MS (Section \ref{subsec:sfrmstar}).} However, some of the other EMPGs show SFRs significantly higher than the $z\sim0$ MS, which means that not every EMPG is a typical SFG at $z\sim0$.

\subsubsection{Comparison with local dwarf galaxies} \label{subsubsec:dwarf}
The \tcr{center} panel of Figure \ref{fig:remstar_feat} \tcg{shows} that some of the EMPGs have values of $r_{\rm e}$ and $M_{*}$ similar to those of dSphs \tcg{and dIrrs}. As we mention in Section \ref{subsec:sfrmstar}, dSphs are totally different from the EMPGs in terms of star-formation activities. \tcg{Although dIrrs show ongoing star formation\tcg{, in} contrast to dSphs, the majority of the EMPGs have sSFR values higher than those of dIrrs (Figure \ref{fig:sfrmstar}).} In Figure \ref{fig:remstar_feat}, we also show that some of the EMPGs are located near GCs. However, GCs not only have already stopped star-formation activities, but also show S{\'e}rsic indices of $n\sim4$ \citep{Ma2015} higher than most of the EMPGs ($n\sim1$; Section \ref{subsec:prop}). 

We conclude that we cannot find a counterpart \tcr{galaxy} satisfying all the properties of the EMPGs.

\subsubsection{\tcr{Comparison with clumps}} \label{subsubsec:clump}
\begin{figure}[t]
    \centering
    \includegraphics[width=8.0cm]{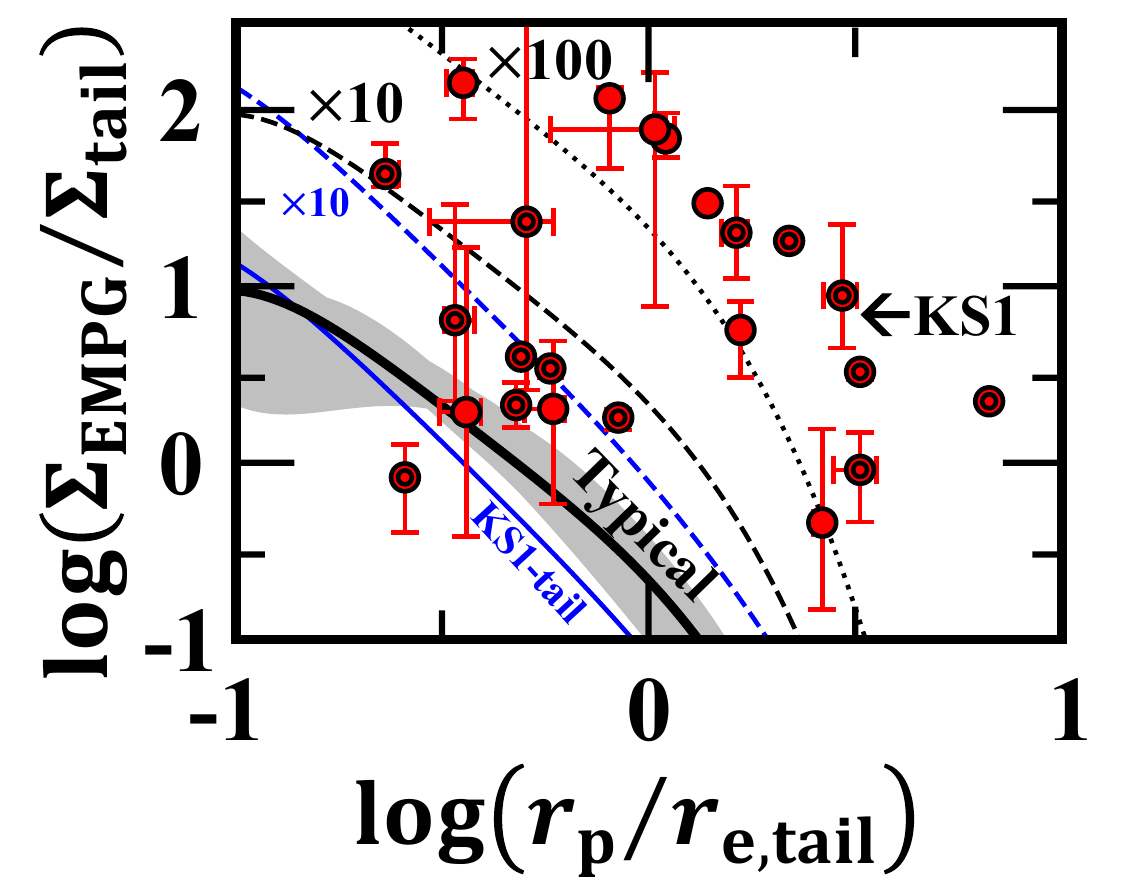}
    \caption{Distribution of $\Sigma_{\rm EMPG}$ and $r_{\rm p}$ normalized by $\Sigma_{\rm tail}$ and $r_{\rm e,\:tail}$, respectively. The black solid curve indicates the normalized SB profile of the typical EMPG-tail. The normalized SB profile of the EMPG-tail varies within the gray shaded region when we change $n$ of the EMPG-tails and $r_{\rm e}$ of the EMPGs within their $\pm68$\% distributions (Section \ref{subsec:prop}). The black dashed and dotted curves show the normalized SB profiles 10 and 100 times larger than that of the typical EMPG-tail. \tcra{The blue solid and dashed curves represent the normalized SB profile of KS1-tail with $n=2.51$ and $\times10$ of the KS1-tail profile, respectively.}}
    \label{fig:ld}
\end{figure}

\tcr{In Section \ref{subsec:remstar}, we point out the size--mass relations of the clumps are similar to those of $z\sim0$ SFGs. Thus, if we use only the $r_{\rm e}$--$M_{*}$ relations, we cannot tell whether the EMPGs are clumps of the EMPG-tails or individual galaxies. \citet{Wuyts2012} report that $z\sim1$--2 clumpy galaxies have clumps whose $V$-band SBs are at most $\sim1$ dex larger than the SB profile of the host galaxies, although such relations have not been investigated for the clumps of \citet{Elmegreen2013}. The SB excesses possibly get smaller in the $i$ band because the $i$-band luminosity is less affected by strong emission lines (Section \ref{subsec:size}) and thus stellar ages. 
If a clump nevertheless has an $i$-band SB more than $\sim1$ dex larger than that of the host galaxy, the clump may be a system that is separate from the host galaxy.
Therefore, estimating how much $i$-band SB of the EMPGs exceed the EMPG-tail SB profile is important to understand whether the EMPGs are likely to be star-forming clumps of the EMPG-tails.}
Here we derive $i$-band SBs of the EMPGs, $\Sigma_{\rm EMPG}$, normalized by $i$-band SBs of the EMPG-tails, $\Sigma_{\rm tail}$. The ratio $\log(\Sigma_{\rm EMPG}/\Sigma_{\rm tail})$ can be calculated by

\begin{equation}
    \tcg{\log(\Sigma_{\rm EMPG}/\Sigma_{\rm tail})=-0.4(\bar{\mu}_{\rm e,\:EMPG}-\bar{\mu}_{\rm e,\:tail}),}
    \label{equ:sb}
\end{equation}
\tcg{where}
\begin{eqnarray}
    \bar{\mu}_{\rm e,\:EMPG}&=&\tcr{M_{i,\:\rm EMPG}}+2.5\log(2\pi r^{2}_{\rm e,\:EMPG}), \nonumber \\
    \bar{\mu}_{\rm e,\:tail}&=&\tcr{M_{i,\:\rm tail}}+2.5\log(2\pi r^{2}_{\rm e,\:tail}),
    \label{equ:meansb}
\end{eqnarray}
\tcr{where $M_{i,\:\rm EMPG}$ and $M_{i,\:\rm tail}$ are absolute magnitudes of the EMPGs and the EMPG-tails, respectively.}
The projected distance $r_{\rm p}$ between each pair of EMPG and EMPG-tail is derived from the best-fit coordinates obtained by the SB S{\'e}rsic profile fitting (Section \ref{subsec:size}). 
Figure \ref{fig:ld} presents the distribution of $\Sigma_{\rm EMPG}/\Sigma_{\rm tail}$ and $r_{\rm p}/r_{\rm e,\:tail}$.
\tcra{For comparison, we derive the normalized SB profile of the EMPG-tails from SB of the 2D S{\'e}rsic profile at a given ($r_{\rm p}/r_{\rm e,\:tail}$) divided by the average SB within $r_{\rm e,\:tail}$. 
The black solid curve represents the normalized SB profile with $n=0.93$ that is the median $n$ of the EMPG-tails. 
The gray shaded region represents how much the SB profile of the EMPG-tails varies when we change $n$ from 0.7 to 1.76. 
The $n$ range corresponds to the $\pm68$\% percentiles of S{\'e}rsic indices of the EMPG-tails, and 19 out of the 23 EMPG-tails have $n$ within the range. 
} 
\tcr{We find 10 out of the 23 EMPGs with the EMPG-tail ($10/23=43$\%) have $\Sigma_{\rm EMPG}/\Sigma_{\rm tail}$ at most $\sim1$ dex larger than the normalized SB profile of the typical EMPG-tail (black solid curve). The SB excesses are comparable to those of the star-forming clumps of \citet{Wuyts2012}. We conclude that 43\% of the EMPGs are likely to be star-forming clumps of the EMPG-tails.}

Conversely, we identify the rest 13 EMPGs with the EMPG-tail ($13/23=57$\%) whose $\Sigma_{\rm EMPG}/\Sigma_{\rm tail}$ are at least $\sim2$ dex larger than the normalized SB profile of the typical EMPG-tail. 
\tcra{Although KS1-tail has a relatively-large $n$ of 2.51, we confirm that $\Sigma_{\rm EMPG}/\Sigma_{\rm tail}$ of KS1 is $>2$ dex larger than the SB profile of KS1-tail (the blue solid curve). }
The 13 EMPGs do not resemble the clumps in the $z\sim1$--2 clumpy galaxies \citep{Wuyts2012}\tcra{, which implies that the 13 EMPGs are a population different from that of clumps in galaxies}. 
These large SB excesses cannot possibly be explained by only ages\tcra{, because starbursts lose their $i$-band luminosity by only $\sim1$ dex for the first 1 Gyr (regarding metal-poor models; \citealt{Leitherer1999}). 
The SBs of the 13 EMPGs cannot be equal to those of the EMPG-tails even after 1 Gyr, unless the EMPGs grow in size by a factor of $\gtrsim3$.}

\subsection{EMPG-tail} \label{subsec:assoc}
\begin{figure}[t]
    \centering
    \includegraphics[width=8.0cm]{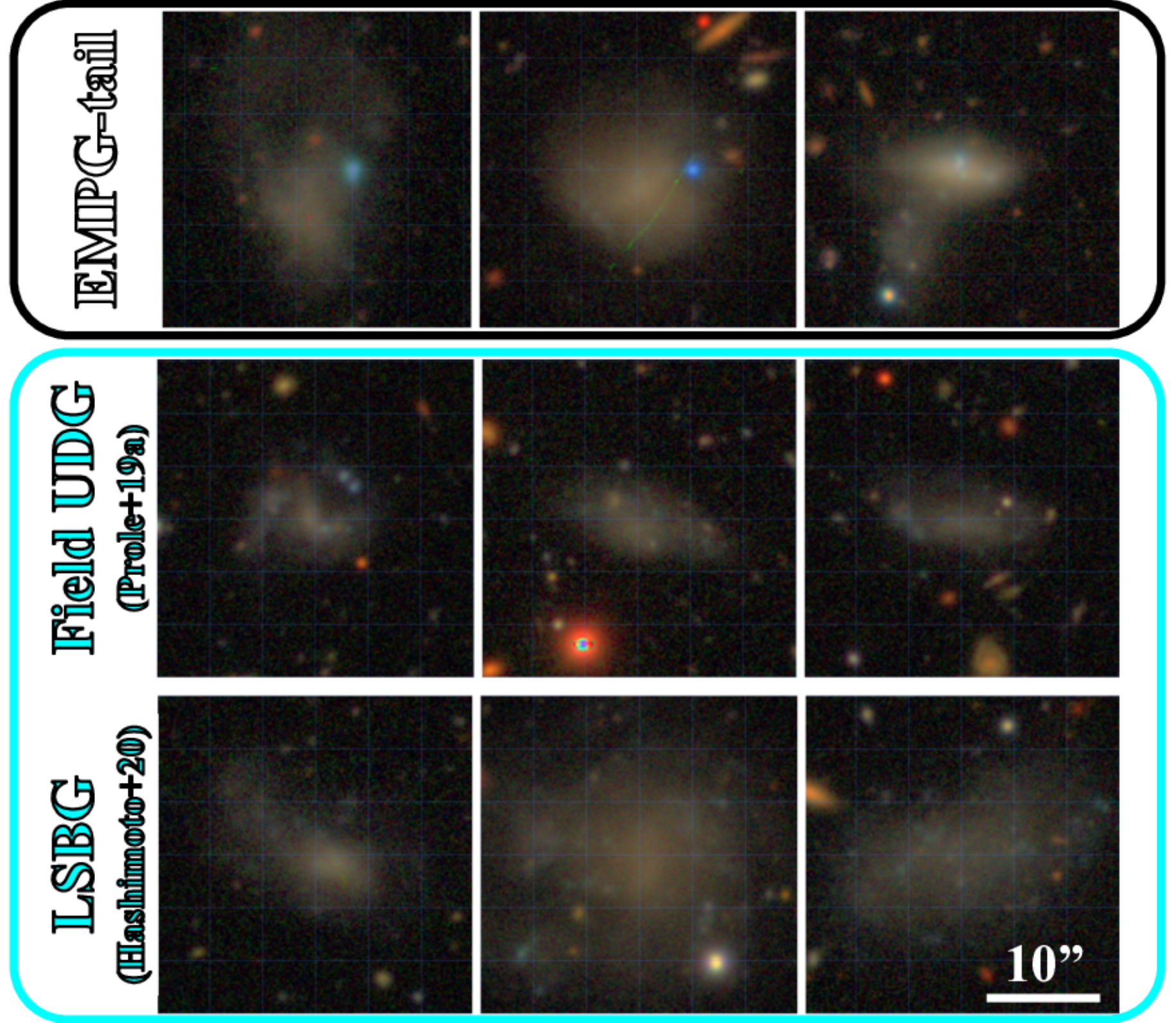}
    \caption{HSC $gri$-composite images of the EMPG-tails (top row), field UDGs (middle row; \citealt{Prole2019}), and LSBGs (bottom row; \citealt{Hashimoto2020}; \citealt{Greco2018}). \tcg{Coordinates of the field UDGs \citep{Prole2019} are provided by D. Prole (in private communication).}}
    \label{fig:ag_udg}
\end{figure}

In Sections \ref{subsec:remstar}, we find that many of the EMPG-tails \tcg{fall around the $r_{\rm e}$--$M_{*}$ relation of} \tcg{dSphs, dIrrs, and} UDGs. \tcg{We can exclude dSphs from a counterpart candidate of the EMPG-tails because most of dSphs are located near the host galaxies (within a virial radius; \citealt{McConnachie2012}), while \tcg{the EMPG-tails are located in an isolated environment (more isolated than typical local galaxies; \tcr{e.g., \citealt{Filho2015};} Paper I).} In contrast, dIrrs are relatively apart from the host galaxies \citep{McConnachie2012}.} UDGs are classified into two groups: those in galaxy clusters \citep{VanDokkum2015} and those in blank fields (field UDGs; \citealt{Prole2019}). Thus, the EMPG-tails may be in \tcg{environments similar to those of} field UDGs. A number of field UDGs reported by \citet{Prole2019} (especially the UDG in the middle left panel of Figure \ref{fig:ag_udg}) have blue star-forming clumps or galaxies, which are also similar to the EMPG-tails. Additionally, typical UDGs have $n<1.5$ \citep{Prole2019}, which are also supported by the result of zoom-in cosmological simulations \citep{DiCintio2017}.

\subsection{Dynamical Relation between EMPG and EMPG-tail} \label{subsec:rel}
\begin{figure*}[t]
    \centering
    \includegraphics[width=18.0cm]{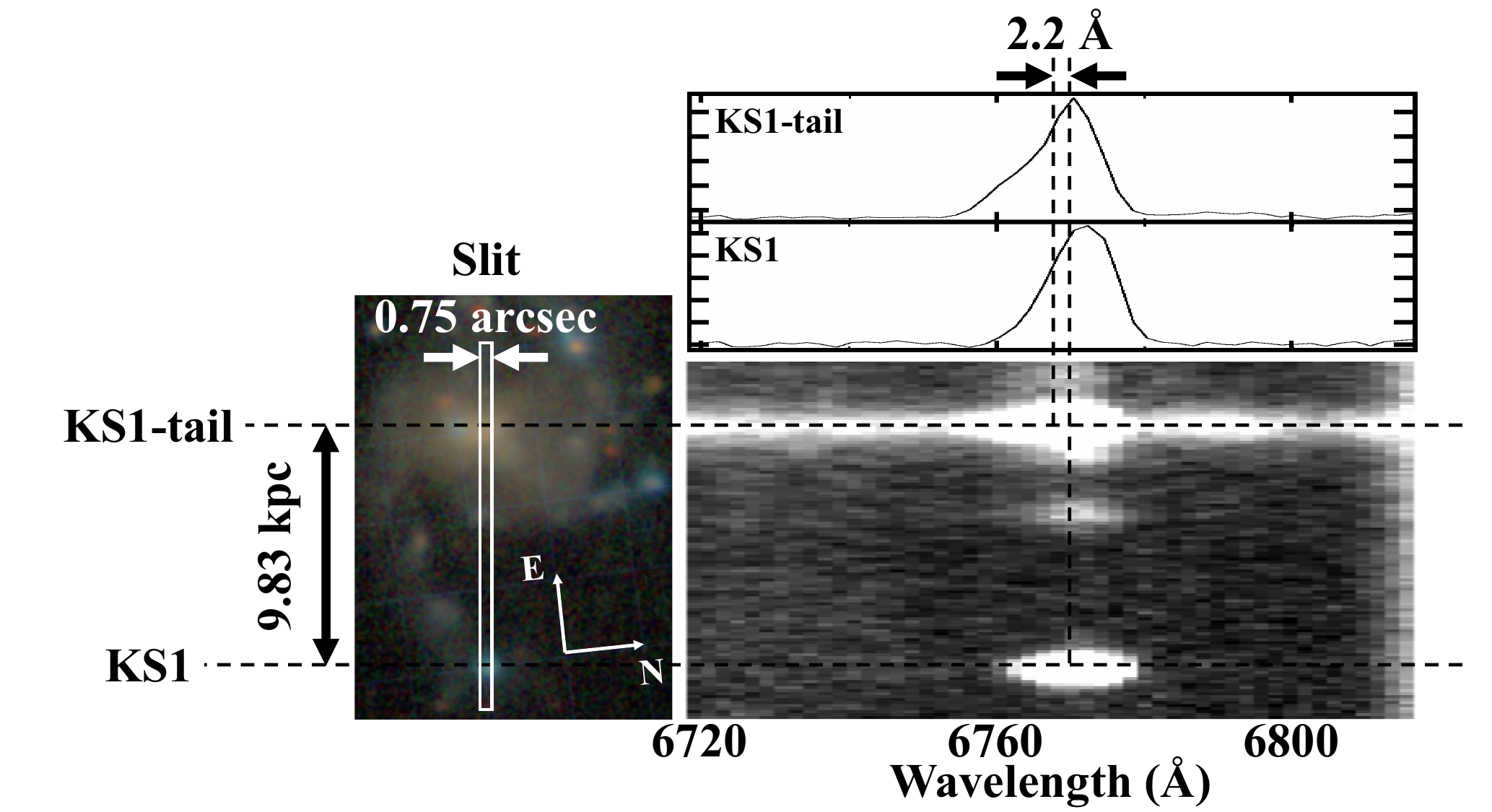}
    \caption{One-dimensional spectra of KS1 and KS1-tail (top right), the $gri$-composite image of KS1 and KS1-tail (bottom left), and two-dimensional spectra of KS1 and KS1-tail (bottom right). The spectra was taken by LDSS-3. The slit width is 0.75 arcsec. \tcr{The vertical dashed lines indicate the barycenters of the spectra of KS1 and KS1-tail.}}
    \label{fig:2d_es1}
\end{figure*}

\begin{figure*}[t]
    \centering
    \includegraphics[width=18.0cm]{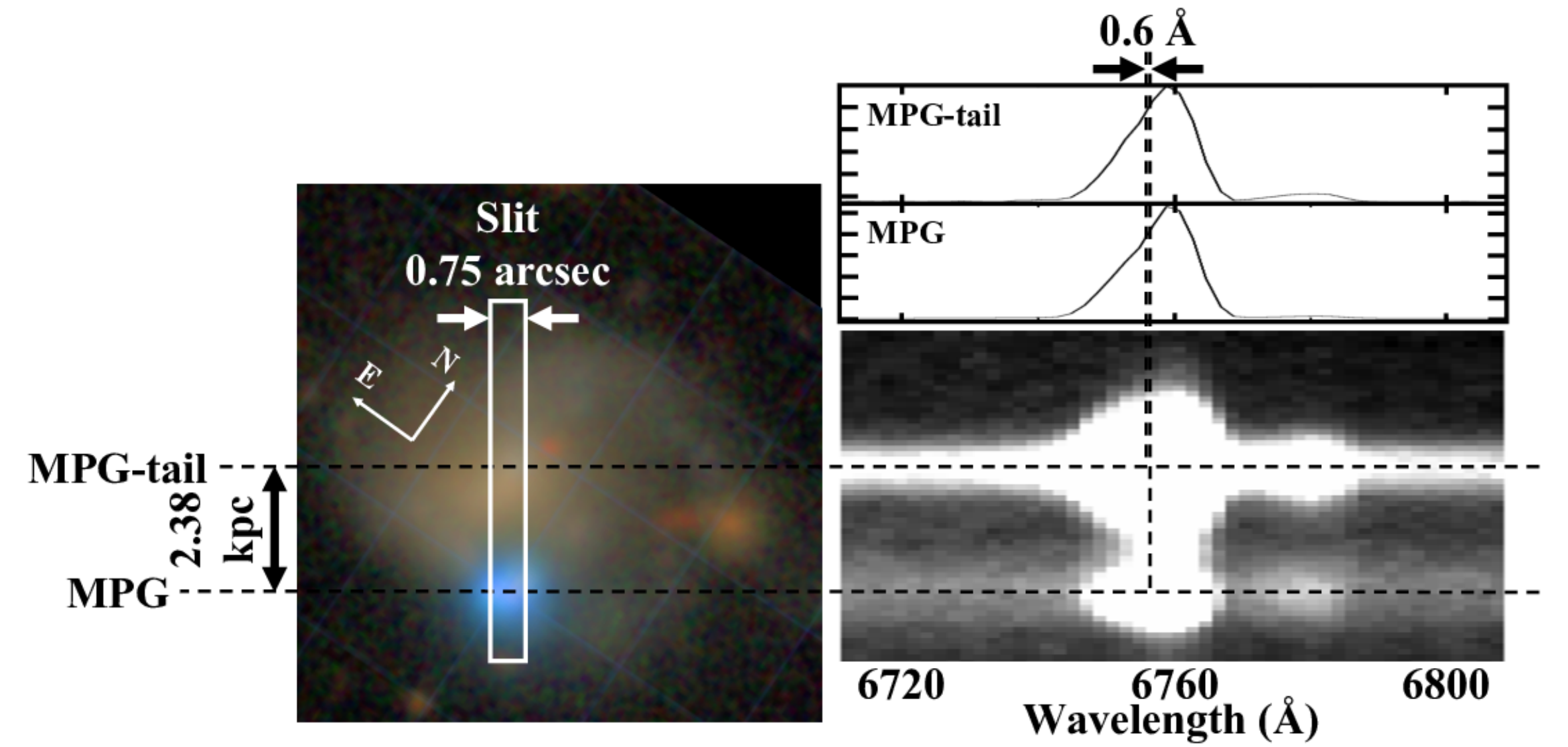}
    \caption{Same as Figure \ref{fig:2d_es1}, but for MPG (see Section \ref{subsec:empg}). Unlike KS1, MPG is located at almost the same redshift as MPG-tail.}
    \label{fig:2d_ms1}
\end{figure*}

\begin{figure}[t]
    \centering
    \includegraphics[width=8.0cm]{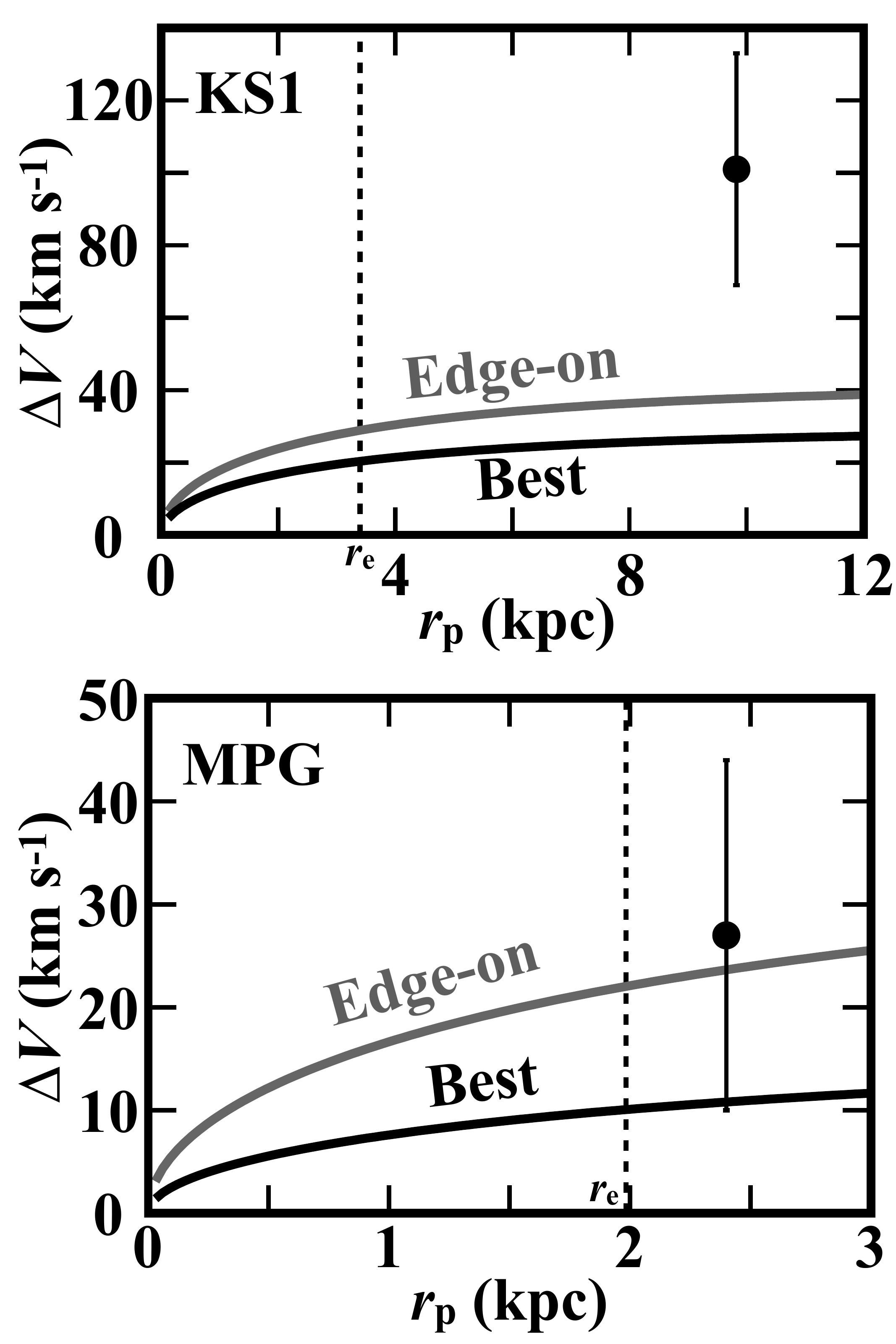}
    \caption{Relation between $\Delta V$ and $r_{\rm p}$ of KS1 (top) and the MPG (bottom). The black circles indicate the observed $\Delta V$ and $r_{\rm p}$ of KS1 and MPG. The black solid lines show the estimated rotation curves of the tails. The gray solid lines represent the maximum $\Delta V$ that the tails can produce, corresponding the case if the tails are edge on. The vertical dotted lines are the $r_{\rm e}$ values of the tails.}
    \label{fig:rot}
\end{figure}

In this section, we discuss dynamical relations between the EMPGs and EMPG-tails. \tcr{In our sample}, KS1-tail is \tcg{the} only EMPG-tail whose spectrum is available, which allows us to evaluate whether KS1 is dynamically related to KS1-tail. \tcr{As shown in Figure \ref{fig:ld}, KS1 has the $i$-band SB more than $\sim2$ dex larger than the normalized SB profile of the typical EMPG-tail, implying that KS1 is a different system that has not been reported yet.} MPG (see Section \ref{subsec:empg}) also has a tail (hereafter MPG-tail) whose spectrum is available. Figure \ref{fig:2d_es1} (Figure \ref{fig:2d_ms1}) presents spectra of KS1 (MPG) and KS1-tail (MPG-tail). 
We measure barycenters of H$\alpha$ emission lines.  
\tcra{To estimate errors of the barycenters, we fluctuate the observed spectrum based on a noise spectrum that contains photon noise. 
We embed the fluctuated spectrum into a continuum spectrum that is randomly selected from the wavelength range of 5100--5800 \AA. 
We repeat the procedure 1000 times for each EMPG or EMPG-tail. 
}
We finally obtain the \tcr{barycenter} differences with the 16th and 84th percentiles of \tcr{$\Delta \lambda=2.2\pm0.7$} and \tcr{$0.6\pm0.4\:{\rm \AA}$} for KS1 and MPG, respectively \tcr{(the right top panels of Figures \ref{fig:2d_es1} and \ref{fig:2d_ms1})}. The \tcr{barycenter} differences correspond to the relative velocities with the 16th and 84th percentiles of \tcr{$\Delta V=101\pm32$} and \tcr{$\Delta V=27\pm17$ km s$^{-1}$}, respectively. Using best-fit coordinates of KS1 and KS1-tail obtained by the SB S{\'e}rsic profile fitting (Section \ref{subsec:size}), the projected distance $r_{\rm p}$ between KS1 and KS1-tail is 9.83 kpc. Similarly, the value of $r_{\rm p}$ between MPG and MPG-tail is 2.40 kpc. For both KS1 and MPG, errors of the $r_{\rm p}$ are smaller than 0.01 kpc. Because $\Delta V$ and $r_{\rm p}$ are smaller than $\sim100\:{\rm km}\:{\rm s}^{-1}$ and $10$ kpc, respectively, KS1 (MPG) \tcg{may} be dynamically related to KS1-tail (MPG-tail). 

In figure \ref{fig:rot}, we plot $\Delta V$ and $r_{\rm p}$ of KS1 (top) and MPG (bottom). Here we investigate whether KS1 (MPG) is the structure on the dynamical system of KS1-tail (MPG-tail). In Section \ref{subsec:prop}, we identify that the EMPG-tails have low S{\'e}rsic indices of $n\sim1$ that is indicative of disk galaxies. Moreover, the images of KS1-tail and MPG-tail show internal structures similar to spiral arms on disk galaxies (the bottom left panels of Figures \ref{fig:2d_es1} and \ref{fig:2d_ms1}). These morphological properties indicate KS1-tail and MPG-tail are probably disk galaxies that are dynamically supported by rotational motions. We thus estimate rotation curves of the EMPG-tails. 

\tcg{Below, we draw rotation curves of KS1-tail and MPG-tail and discuss how KS1 and MPG are dynamically associated with KS1-tail and MPG-tail, respectively.} Because $r_{\rm p}$ of KS1 (MPG) is larger than $r_{\rm e}$ of KS1-tail (MPG-tail), dynamics around KS1 (MPG) is dominated by the dark-matter (DM) halo of KS1-tail (MPG-tail). We assume the density profile of the DM halo of the EMPG-tail as a \tcg{Navarro-Frenk-White (hereafter, NFW)} profile derived with CDM models \citep{Navarro1996}. The circular velocity of the NFW halo $V_{\rm c}(r)$ can be calculated by

\begin{equation}
    \left(\frac{V_{\rm c}(r)}{V_{200}}\right)^{2}=\frac{1}{r/r_{200}}\frac{\ln[1+C(r/r_{200})]-\frac{C(r/r_{200})}{1+C(r/r_{200})}}{\ln(1+C)-\frac{C}{1+C}},
    \label{equ:rot}
\end{equation}
where $V_{200}$, $r$, and $r_{200}$ represent the virial velocity, the radius, and the virial radius, respectively. The parameter $C$ describes the concentration, which is roughly correlated with the surface brightness \citep{Navarro1998}. We assume $C=5$ for galaxies with low surface brightnesses \citep{Navarro1998}. The values of $V_{200}$ and $r_{200}$ in units of ${\rm km}\:{\rm s}^{-1}$ and kpc are described with

\begin{equation}
    V_{200}=\left(\frac{M_{200}}{2.33\times10^{5}\:{\rm M}_{\odot}}\right)^{1/3}
    \label{equ:virv}
\end{equation}
and

\begin{equation}
    r_{200}=\frac{GM_{200}}{V_{200}^{2}},
    \label{equ:virr}
\end{equation}
respectively\tcr{, where $G$ is the gravitational constant of $6.67\times10^{-8}$ cm$^{3}$ s$^{-2}$ g$^{-1}$}. Here $M_{200}$ is the DM halo mass. We obtain the relations between $M_{*}$ and $M_{200}$\tcr{, both in units of the stellar mass,} assuming the stellar-to-halo mass relations for low-mass galaxies \citep{Brook2014}

\begin{equation}
    M_{*}=\left(\frac{M_{200}}{79.6\times10^{6}\:{\rm M}_{\odot}}\right)^{3.1}
    \label{equ:Bro14}
\end{equation}
that is applicable for UDGs and LSBGs \citep{Prole2019b}. The virial velocity and the virial radius of KS1-tail (MPG-tail) are estimated to be $V_{200}=51.8$ ($45.7\:{\rm km}\:{\rm s}^{-1}$) and $r_{200}=51.9$ ($45.7\:{\rm kpc}$), respectively. Comparing \tcg{the $r_{\rm p}$ values} between KS1 (MPG) and KS1-tail (MPG-tail), we find that KS1 (MPG) is located within the virial radius of KS1-tail (MPG-tail). Then we estimate the inclinations $i$ of the EMPG-tails using the relation of

\begin{equation}
    {\rm cos}^{2}i=\frac{q^{2}-q_{0}^{2}}{1-q_{0}^{2}},
    \label{equ:Fou90}
\end{equation}
where $q$ and $q_{0}$ are the axis ratios of the EMPG-tail with arbitrary $i$ and $i=90^{\circ}$, respectively. We adopt $q_{0}=0.3$ that is applicable for disk galaxies \citep{Fouque1990}. KS1-tail and MPG-tail have $q$ of 0.74 and 0.90, respectively. 
Substituting $q$ in Equation \ref{equ:Fou90}, we obtain $i=45^{\circ}$ and $i=27^{\circ}$, respectively. 

Using Equations \ref{equ:rot} and \ref{equ:Fou90}, we calculate the rotation curve along the line of sight $\Delta V(r)=V_{\rm c}(r)\sin{i}$ from $M_{*}$. The black curves in Figure \ref{fig:rot} show the rotation curves of KS1-tail and MPG-tail. The gray curves indicate the maximum velocity cases corresponding to the edge-on ($i=90^{\circ}$) cases of KS1-tail and MPG-tail. 
We find that $\Delta V$ between KS1 and KS1-tail is significantly higher than $\Delta V$ expected by the rotation curve even in the maximum velocity case \tcr{($\Delta V\sim40$ km s$^{-1}$ at $r_{\rm p}\sim10$ kpc)}. 
\tcra{This velocity excess indicates that KS1 is dynamically independent of a disk structure of KS1-tail. Such a velocity excess is not be seen in previously known tadpole galaxies (\citealt{SanchezAlmeida2013}; \citealt{Olmo-Garcia2017}; see Section \ref{sec:intro}).}
At the same time, we find MPG whose $\Delta V$ is comparable to the rotation velocity (Figure \ref{fig:rot} bottom), indicative that MPG is a system similar to the tadpole galaxies. Thus we do not think the methodological differences cause the discrepancy between KS1 and the tadpole galaxies.

The bottom left panel of Figure \ref{fig:2d_es1} indicates that \tcra{
KS1 and KS1-tail are connected with a filamentary structure.
One possible scenario is that KS1 is located in the filamentary structure with a large proper motion. 
However, KS1 is unlikely a clump in KS1-tail due to the large SB difference with respect to KS1-tail as shown in Section \ref{subsubsec:clump}. 
The filamentary structure may be a gas stream now accreting onto KS1-tail as shown in Figure 1 of \citet{Tumlinson2017}\footnote{Because the figure of \citet{Tumlinson2017} is a schematic painting, we note that some of expressions are exaggerated such as outflows.}, or a tidal tail created by gravitational interactions between KS1 and KS1-tail.
In either case, we are likely to identify the metal-poor star-forming system just now infalling into the EMPG-tail, which supports the idea more directly that matters of the extremely metal-poor starbursts come from outside of the EMPG-tails (\citealt{SanchezAlmeida2015}; see Section \ref{sec:intro}).
}

\tcra{There is also a possibility that some other EMPGs, especially 57\% of the EMPGs with the large SB differences like KS1 (Section \ref{subsubsec:clump}), have large velocity excesses with respect to the EMPG-tails. 
To investigate such EMPGs, we need long-slit or integral-field spectroscopy for EMPGs. }

\section{Summary} \label{sec:sum}
We present the morphology and stellar population of 27 EMPGs. We conduct multi-component SB profile fitting for the HSC $i$-band images of the EMPGs with the {\sc Galfit} software, carefully removing the SB contributions of EMPG-tails. The major results of our study are summarized below. 

\begin{enumerate}
    \item The EMPGs have a median S{\'e}rsic index of \tcg{$n=1.1$} and a median effective radius of \tcr{$r_{\rm e}=200$ pc}, suggesting that typical EMPGs \tcg{have} very compact disks. We estimate a median stellar mass of the EMPGs to be a small value of \tcra{$\log(M_{*}/{\rm M}_{\odot})=6.0$}. 
    \item We compare our galaxies with $z\sim 6$ galaxies and local galaxies on the size-mass ($r_{\rm e}$--$M_*$) diagram. \tcg{The majority of our galaxies obey} a $r_{\rm e}$--$M_*$ relation similar to $z\sim0$ star-forming galaxies rather than $z\sim6$ galaxies. \tcg{Most low-$z$ EMPGs do not seem to be analogs of $z\sim6$ galaxies.} 
    \item \tcg{Twenty-three out of the 27} EMPGs show detectable EMPG-tails within \tcg{a} projected distance of 10 kpc. The EMPG-tails have median values of $n=0.9$, \tcr{$r_{\rm e}=1.6$ kpc, and $\log(M_{*}/{\rm M}_{\odot})=7.5$} that are similar to those of local \tcg{dIrrs and} UDGs. 
    \item \tcr{We find that many of the EMPGs have $r_{\rm e}$--$M_{*}$ relations similar to those of star-forming clumps. Calculating $i$-band SB excesses of the EMPGs with respect to the EMPG-tail profile, we estimate that 43\% of the EMPGs are likely to be star-forming clumps of the EMPG-tails.}
    \item The spectrum of one pair of EMPG and EMPG-tail, so far available, indicate that the EMPG-tail is dynamically related to the EMPG with \tcg{a} median velocity difference of \tcr{$\Delta V=101$ km s$^{-1}$}. This moderately-large $\Delta V$ cannot be explained by the dynamics of the EMPG-tail, but \tcg{is likely due to} infall on the EMPG-tail. 
\end{enumerate}

\acknowledgments

We thank Koki Kakiichi and Chengze Liu for having useful discussions. We are grateful to Daniel Prole for providing the coordinates of UDGs in \citet{Prole2019}. \tcg{This paper includes data gathered with the 6.5 meter Magellan Telescopes located at Las Campanas Observatory, Chile. We thank the staff of Las Campanas for their help with the observations.} The Hyper Suprime-Cam (HSC) collaboration includes the astronomical communities of Japan and Taiwan, and Princeton University. The HSC instrumentation and software were developed by the National Astronomical Observatory of Japan (NAOJ), the Kavli Institute for the Physics and Mathematics of the Universe (Kavli IPMU), the University of Tokyo, the High Energy Accelerator Research Organization (KEK), the Academia Sinica Institute for Astronomy and Astrophysics in Taiwan (ASIAA), and Princeton University. Based on data collected at the Subaru Telescope and retrieved from the HSC data archive system, which is operated by Subaru Telescope and Astronomy Data Center at NAOJ. This work was supported by the joint research program of the Institute for Cosmic Ray Research (ICRR), University of Tokyo. The Cosmic Dawn Center is funded by the Danish National Research Foundation under grant No. 140. 
S.F. acknowledges support from the European Research Council (ERC) Consolidator Grant funding scheme (project ConTExt, grant No. 648179). 
This project has received funding from the European Union's Horizon 2020 research and innovation program under the Marie Sklodowska-Curie grant agreement No. 847523 `INTERACTIONS'.
This work is supported by World Premier International Research Center Initiative (WPI Initiative), MEXT, Japan, as well as KAKENHI Grant-in-Aid for Scientific Research (A) (15H02064, 17H01110, and 17H01114) through Japan Society for the Promotion of Science (JSPS). Takashi Kojima, Kohei Hayashi, Ken Mawatari, Masato Onodera, Yuma Sugahara, and Kiyoto Yabe are supported by JSPS KAKENHI Grant Numbers, 18J12840, 18J00277, 20K14516, 17K14257, 18J12727, and 18K13578, respectively.

%

\bibliography{library}


\end{document}